\documentclass[a4paper, notitlepage,12pt]{article}
\usepackage[utf8]{inputenc}
\usepackage{graphicx}
\usepackage{xcolor}
\usepackage{appendix}
\usepackage{natbib}

\usepackage[left=1in, right=1in, top=1in, bottom=1in]{geometry}

\usepackage{placeins}
\usepackage{booktabs}
\usepackage{hyperref}
\usepackage{setspace}
\usepackage{subfiles}
\usepackage{lscape}

\usepackage{amssymb}
\usepackage{amsfonts}
\usepackage{amsmath}
\usepackage{amsthm}
\usepackage{caption,subcaption}

\usepackage{soul}
\definecolor{highlightcolor}{RGB}{255,255,150}
\sethlcolor{highlightcolor}

\newtheorem{theorem}{Theorem}

\title{Saving for sunny days: The impact of climate (change) on consumer prices in the euro area}
\author{Paulo M. M. Rodrigues\thanks{Banco de Portugal and Nova School of Business and Economics, pmrodrigues@bportugal.pt}, Mirjam Salish\thanks{Oesterreichische Nationalbank, mirjam.salish@oenb.at}, Nazarii Salish\thanks{Universidad Carlos III de Madrid, nsalish@eco.uc3m.es \newline
Disclaimer: Opinions expressed by the authors do not necessarily reflect the official viewpoint of the Banco de Portugal, the OeNB or the Eurosystem. }}
\date{Preliminary Version: January 2024 \\ Do not cite without permission}

\begin{document}

\maketitle

\begin{abstract}
\begin{singlespace}
Climate (change) affects the prices of goods and services in different countries or regions differently. Simply relying on aggregate measures or summary statistics, such as the impact of average country temperature changes on HICP headline inflation, conceals a large heterogeneity across (sub-)sectors of the economy. Additionally, the impact of a weather anomaly on consumer prices depends not only on its sign and magnitude, but also on its location and the size of the area affected by the shock. This is especially true for larger countries or regions with diverse climate zones, since the geographical distribution of climatic effects plays a role in shaping economic outcomes. Using time series data of geolocations, we demonstrate that relying solely on country averages fails to adequately capture and explain the influence of weather on consumer prices in the euro area. We conclude that the information content hidden in rich and complex surface data can provide valuable insights into the role of weather and climate variables for price stability, and more generally may help to inform economic policy. 
\end{singlespace}

\noindent \textbf{Keywords:} Inflation, climate change, surface time-series data, heterogeneity, common factors, functional data analysis

\noindent \textbf{JEL codes:} C18, E31, E37, Q54
\end{abstract}

\newpage

\section{Introduction}
\setstretch{1.5}
Climate change and the dependence on fossil fuels are one of the greatest challenges of our time. Anthropogenic greenhouse gas emissions contribute to a rise in global temperatures and are linked to the increasing incidence of extreme weather events such as droughts, floods, wildfires, and heat waves. Consequences of global warming include rising sea levels, melting of glaciers, reduction of biodiversity, as well as heat-related increases in morbidity and mortality rates, all of which are interrelated. Thus, climate change impacts economic activity in a variety of ways: From a rise in the volatility of agricultural output, over productivity losses, supply and demand imbalances, the destruction of infrastructure and global supply chains, increased uncertainty, to structural and sectoral shifts.

The (observed and expected future) consequences for economic activity are so severe that climate protection is no longer seen as the exclusive responsibility of non-governmental organizations (NGOs) or environmentalists. The network of greening the financial system has grown from eight (founding) members in 2017 to more than 100 members as of December 2021. In 2021 the European Central Bank (ECB) vouched to support the European Union's (EU) plans for climate protection in its monetary policy strategy as long as this would not be at odds with its primary mandate of price stability. Several other central banks also take sustainability considerations into account. Consequently, one key question from a central bank’s perspective is to understand the relationship between climate change and price stability. This is becoming ever more important in light of the recent high inflation period. Not only central banks, but also national governments need to understand this relation. Price stability is the main priority of central banks around the globe for a reason: Sustainable and inclusive growth is difficult to achieve without stable prices and it is not only central banks that have to answer for high inflation rates.  To be able to take evidence-based decisions and design policies that protect the most vulnerable members of society while adapting to climate change, it is crucial to understand the implications of climate change on price developments.

Climate can impact consumer prices directly and indirectly through various channels. Extreme weather events and natural hazards, such as droughts, wildfires or floods destroy harvests, infrastructure, and livestock. Heat waves, long and cold winters or shifts in precipitation patterns may raise the costs of agricultural production and increase the volatility of output, lower labor productivity and exacerbate inequality. At the same time, climate protection policies have direct and indirect effects on prices. The price of carbon, as measured by the EU's emissions trading system (EU ETS) allowance prices, increased markedly throughout 2021, and in many EU countries national policies complementary to the EU ETS directly raise the prices for fossil energy sources. The direct inflationary effects of mitigation policies are often easier to measure and quantify than the effects of climate change itself leading to a possible underestimation of the threat that climate change poses to price stability.

The existing literature on the impact of climate (change) on inflation (see Section \ref{LitRev}) so far has mainly focused on extreme temperatures (\citealp{FacciaParkerStracca2021}), natural disasters (\citealp{DafermosKriwoluzkyVargasVolzWittich2021} and \citealp{Parker2018}) and precipitation (\citealp{Moessner2022} and \citealp{KotzKuikLisNickel2023}). However, climate change has many dimensions and to fully understand its impact on prices, studying each of them in isolation or only focusing on a particular subset may be insufficient and lead to biased estimates (\citealp{AuffhammerSolomonHsiangSchlenkerSobel2013}). For instance, low wind speeds (as experienced in Europe in the summer of 2021), reduce wind energy and therefore impact electricity consumer prices as well as the prices for the fossil fuels needed as substitutes.\footnote{There is no widespread agreement regarding the relation between climate change and wind speeds. However, IPCC (2021) estimates that climate change will reduce average wind speeds in Europe by up to 10\%.} Similarly, insufficient rainfall increases the costs of agricultural production and reduces water quantities and speed in rivers. This in turn may dampen hydro-power based electricity generation. Heavy rainfall, on the other hand, may damage crops, lead to flooding and increase landslide risk. Additionally, extreme precipitation can degrade water quality by overwhelming wastewater systems or by carrying pollutants to lakes, streams or bays and thereby harm ecosystems and increase the risk for waterborne diseases. Changes in solar radiation, often referred to as “solar brightening” (experienced particularly in North America and Europe over the last decades) and “solar dimming” (which in recent years concerned particularly China and India) impacts crop yields, where increased brightening has been shown to have contributed to surges in crop yield growth (\citealp{TollenaarFridgenTyagiStackhouseKumudini2017} and \citealp{Proctor2021}). Temperatures affect the demand for energy needed for heating or cooling and thereby may drive up consumer prices for energy. Energy and food are both essential inputs in the production of various goods and services, and any upward shifts in their prices can potentially trigger cascading effects.

Even if ambitious climate policies manage to keep global warming below 1.5$^{\circ}$C until the end of the century (which seems an overly optimistic assumption at the current juncture), we need to understand and adapt to the damage already caused. To this end, it is important to understand the direct short-term as well as potential medium to long-term effects of the different dimensions of climate change on prices. The contribution of this paper is twofold. First, we contribute to the empirical discussion on the impact of weather shocks on consumer price inflation; and second, we show how to use the extensive information of time-series data for geolocations to study the impact of climate change on inflation. 

For the first part, we set out to answer the following questions: What is the  impact of weather anomalies on consumer prices in the euro area and which sectors of the economy are affected most? What is the delay and duration of the impact? What kind of weather events have the strongest impact on inflation and how do different weather events interact with each other? How important are non-linearities regarding the size, sign, timing, location and size of the area of the shock? How heterogeneous is the impact across countries and are there spillovers from one region to another? In addition to including a larger variety of climate-related indicators, we also study the effects at a particularly disaggregated level. \citet{FacciaParkerStracca2021} argue that aggregation across time results in a loss of information. To limit the amount of information loss we use monthly price data for the 20 euro area countries including consumer prices as measured by the harmonised index of consumer prices (HICP) at the European classification of individual consumption according to purpose (ECOICOP) 4 digit level\footnote{\url{https://ec.europa.eu/eurostat/statistics-explained/index.php?title=Glossary:Classification_of_individual_consumption_by_purpose_(COICOP)}}. Climate data is taken from the Copernicus database\footnote{\url{https://www.copernicus.eu/en/access-data}} and includes detailed data on temperatures, precipitation, solar radiation, and wind speed. 

In a first step, we estimate the impact of different weather variables (using monthly country averages) on sectoral consumer price inflation via local projections. The impulse response functions of a change in the deviation of different weather variables from their historical means on subsector inflation rates, show that there is substantial heterogeneity across sectors, countries and timing. While temperature has the largest impact, other climate variables matter as well - particularly at the subsector level. We also confirm the presence of non-linearities by showing that the magnitude and the sign of the weather shocks matter, and again these effects are highly heterogeneous across subsectors and countries. Seasons crucially shape the (dis-)inflationary impacts of weather shocks. While hot summers exert significant upward price pressures in the food sector, a mild autumn or winter has disinflationary effects on energy prices. Besides this, we find that the results crucially depend on the choice of the control variables included in the analysis.

In the second step of our study, we expand our analysis by replacing country-level climate data averages with surface climate data. Conventional tools typically utilised for climate data processing can be overly restrictive, as they do not adequately capture the intricate nature of these datasets, leading often to information loss due to aggregation or other simplifications. By employing novel techniques from functional data analysis we can treat climate data, collected sequentially over time, as a surface (functional) time series with a predefined geographical domain. Furthermore, we propose a new statistical methodology to study potential relationships with such complex data sets and extract common factors. Our method is based on the adaptation of conventional canonical correlation analysis  (CCA) to high-dimensional and infinitely-dimensional settings. 

The remainder of the paper is structured as follows. Section \ref{LitRev} reviews the related literature. In section \ref{data} the data are described. The results for impact of weather shocks on sectoral inflation in a low-dimensional setting are summarized in section \ref{LP}. Section \ref{Associated_factors} describes how to relate high and infinite-dimensional data sets and estimate the impacts of weather shocks using surface-time series data. Finally, section \ref{conclusion} concludes. All proofs are relegated to the appendix.

\section{Literature review \label{LitRev}}

Studies on the economic consequences of climate change frequently focus on the impact of natural disasters on economic growth and are often limited to the “Big Four”: floods, droughts, windstorms and earthquakes; see e.g. \citet{FelbermayrGröschl2014} and \citet{Kousky2014}. Natural disasters tend to reduce economic activity, particularly in developing economies. They may cause negative supply-side shocks by destroying harvests and infrastructure (\citealp{Batten2020ClimateCM} and \citealp{Simola2020}) and consequently raise the costs of production and/or imports. Reconstruction efforts may stimulate demand and thereby raise the prices of reconstruction goods. At the same time, the destruction of wealth may dampen inflation through its negative impact on consumption and investment. Weather shocks can affect global food commodity prices (see e.g. Marini, 2020 or \citealp{TaskinCagliMandaci2021}) and thereby not only impact economic growth but also increase the risk of conflict (see e.g. \citealp{DeWinnePeersman2021a, DeWinnePeersman2021b}). It is again the developing countries that are affected most (\citealp{CiccarelliMarotta2021}). Similarly, \citet{Dell_Jones_Olken_2012} find that higher temperatures reduce economic growth particularly in poor countries. Not only growth rates, but also the level of output can be affected. Additionally, higher 
temperatures lower agricultural output, industrial output, and may negatively affect political stability. Beyond the impact of changes in average temperatures, empirical evidence suggests that weather anomalies (\citealp{FelbermayrGröschlSandersSchippersSteinwachs2022}) or extreme precipitation (\citealp{KotzLevermannWenz2022}) dampen economic growth, and that most of these impacts are likely non-linear (IPCC, 2022b). This is in contrast to the findings of \citet{Deschenes_Greenstone_2007}, who find variation in temperature and precipitation will increase agricultural profits.

Empirical evidence on the impact of weather or climate change on consumer prices is currently an active field of research. \citet{MukherjeeOuattara2021} use a panel-VAR with fixed-effects to show that temperature shocks lead to inflationary pressures. These effects can persist several years after the shock, particularly in developing countries. \citet{DafermosKriwoluzkyVargasVolzWittich2021} provide empirical evidence that natural disasters lead to increases in headline and core inflation in the euro zone. Overall, the effects are small but significant, with food and beverages showing the highest price increases. Moreover, the authors find substantial heterogeneity across the euro zone. More recently, a few studies looked at the impact of extreme temperatures on inflation. \citet{FacciaParkerStracca2021} find that high temperatures tend to increase inflation in the short-run and decrease it in the medium term. Emerging economies are affected more than advanced economies. Hot summers show the largest and most persistent impact on inflation, particularly on the prices for food. Additionally, the effect of climate change on prices is non-linear. Recently \citet{CiccarelliKuikHernández2023} analyse the impact of weather shocks on inflation components in the four largest euro area economies combining high-frequency weather data with monthly inflation. The results suggest significant country asymmetries and seasonal responses of inflation to temperature shocks, mainly via food, energy, and service prices. An increase in monthly mean temperatures has inflationary effects in summer and autumn, with a stronger response in warmer euro area countries. An increase in temperature variability has significant upward inflationary impacts. In both \citet{CiccarelliKuikHernández2023} and \citet{FacciaParkerStracca2021}, weather shocks are defined as deviations from historical averages. \citet{Natoli2023} proposes a different way to construct temperature shocks using county-level data for the US taking into account temperature shocks that agents anticipate based on past temperature realizations. That is, temperature beliefs reflect some learning behavior. In this framework, temperature surprises dampen aggregate demand by reducing GDP, consumer prices and interest rates. 

Only few papers consider weather variables other than temperatures. One of these is \citet{Moessner2022} who use dynamic panel data methods for the estimation of food CPI Phillips curves augmented with climate variables. Precipitation has significant non-linear effects on food inflation with inflation increasing as precipitation becomes either very low or very high. Moreover, temperature has no additional explanatory power for food CPI inflation. \citet{KotzKuikLisNickel2023} also consider temperature and precipitation by calculating so called Standardised Precipitation Evapotranspiration Indices (SPEI) that model dry and wet conditions. Additionally, they combine their results with projections from physical climate models to estimate the inflationary impacts caused by future warming. They identify deviations of monthly mean temperature and daily temperature variability, as well as excess wet and drought conditions and extreme daily rainfall as the main impact channels. Taking into account different weather variables is important to obtain unbiased estimates. For example, temperature and precipitation are historically correlated and this correlation may be location specific (see \citealp{AuffhammerSolomonHsiangSchlenkerSobel2013}).  

In the majority of the papers mentioned above weather variables are aggregated across space (and often time). Weather data, however, is available at a high resolution (and high frequency) and falls into the category of “big data”. As we propose a method for leveraging the information from this data, our paper is also related to the field of big data or high-dimensional data analysis with a particular focus on the use of surface time series data. To describe and model the relationship between sectoral inflation rates and weather, we use techniques from canonical correlation analysis (CCA). CCA is a family of methods that is commonly used to measure the linear relationship between two groups of multidimensional variables (see e.g. \citet{Hotelling1936} or \citet{Hardle_Simar_2007}). In high-dimensional settings, identifying the canonical variables is impossible, when all dimensions are large (see \citet{Bykhovskaya_Gorin_2023}). Therefore, in applications with high-dimensional data, a regularized modification of CCA is commonly used (see e.g. \citet{Tuzhilina_Tozzi_Hastie_2020}). 

\section{Data \label{data}}

\begin{figure}[htp!]                 
  \centering
  \includegraphics[width=12cm]{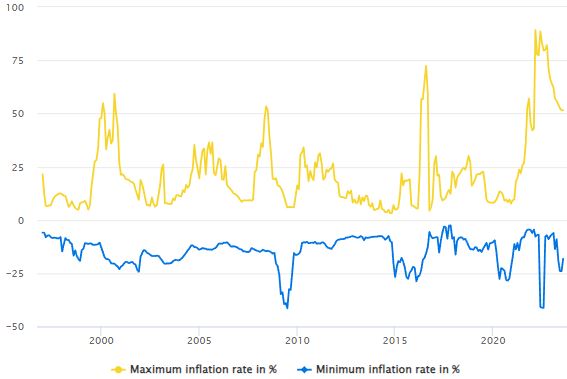}
  \caption{Range of inflation rates in the euro area: The blue line shows the lowest inflation rate and the yellow line the highest inflation rate across 89 subsectors in a given month.}
  \label{fig_inflation}
\end{figure}

Data on consumer prices is taken from Eurostat and includes monthly inflation rates of the HICP starting from January 2001 until December 2021 for 89 (sub)sectors. The items in the HICP are classified according to ECOICOP. We focus on the four main aggregates (energy, food, services and non-energy industrial goods), and their subsectors at ECOICOP 3 and 4 digit level. As can be seen in Figure \ref{fig_inflation}, inflation rates (measured as year-on-year changes in percent) in the euro area varied considerably across the 89 sectors. 

\FloatBarrier

Climate data is taken from the Europe-wide E-OBS\footnote{We acknowledge the E-OBS data set from the EU-FP6 project UERRA (\url{https://www.uerra.eu}) and the Copernicus Climate Change Service, and the data providers in the ECA\&D project (\url{https://www.ecad.eu} and \url{https://surfobs.climate.copernicus.eu/dataaccess/access_eobs.php}).} dataset which consists of daily interpolations of mean temperature, minimum temperature, maximum temperature, precipitation, solar radiation, wind speed, humidity and mean sea-level pressure. Additionally, various other indicators and indices, such as e.g., the number of tropical nights or heating degree days, are available. For the subsequent analysis we include daytime temperatures (average, minimum and maximum), precipitation, wind speed and solar radiation. The E-OBS data set is an ensemble data set, i.e., a climate data set that consists of a number of equally probable realizations. The mean across the members is calculated and is provided as the ``best-guess" fields, which is the data we use in our analysis. The data covers the period back to 1950 (except for wind speed which starts in 1980) and is available as gridded fields at a spacing of either $0.1 \times 0.1$ or $0.25 \times 0.25$ in regular latitude/ longitude coordinates. We use the $0.25 \times 0.25$ spacing which is roughly equivalent to $25km \times 25km$. The usual approach in the production of a gridded data set is to determine the most likely values to interpolate station values to values on a regular grid. Data is available for entire Europe although station density varies across time and geographical locations. No homogeneity corrections are applied to the station data. For more detailed information on the construction of the ensemble data set see \citeauthor{CornesRichardSchrierBesselaarPhilip2018} (2018). We use the period from 1950-1980 as a reference period and calculate the monthly mean for each month of the year based on the daily mean weather variables of the E-OBS data set. Then we compare these values to our period of interest 2001-2021.\footnote{Appendix A provides a description of the steps involved in obtaining the underlying climate data set for the subsequent analysis.} 

\begin{figure}[htp!]                 
  \centering
  \includegraphics[width=16.5cm]{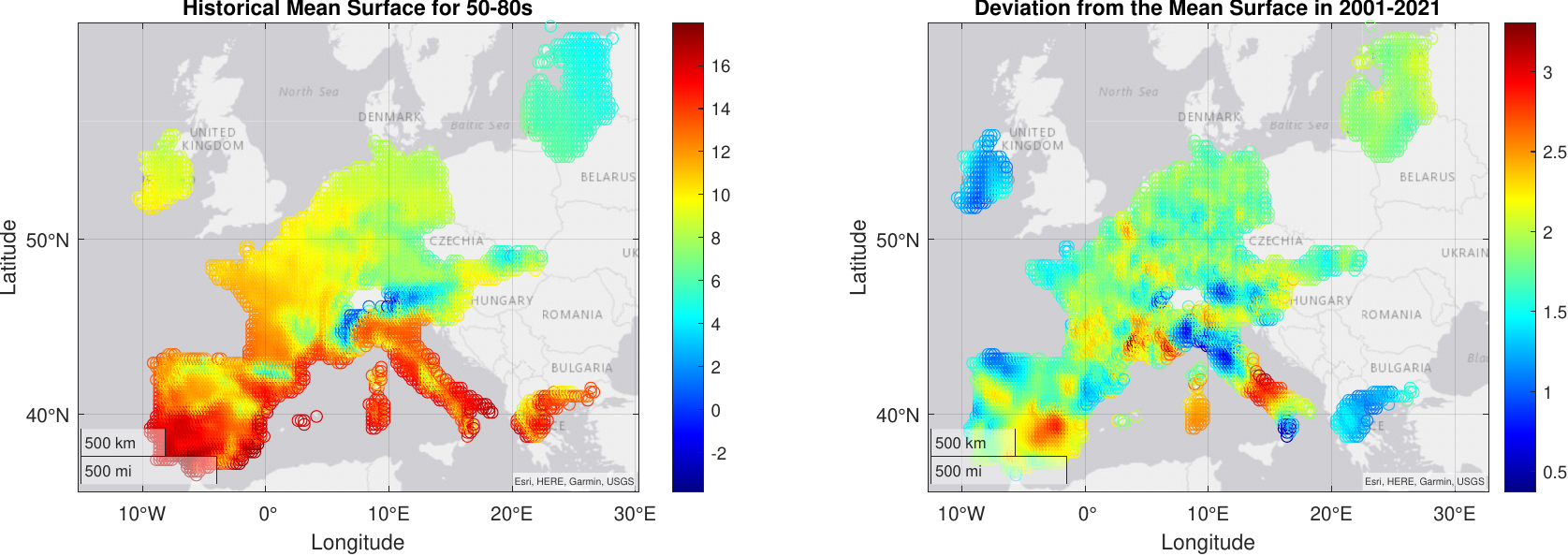}
  \caption{Left Panel: the surface of historical average temperature for the period from 1950 to 1980; Right Panel: average deviation from the historical mean surface for the period from 2001-2021}
  \label{fig:DevMaps}
\end{figure}

By averaging over space we can neither capture whether different regions are affected differently by weather shocks, nor can we understand whether these differences matter for consumer prices - which would be important to quantify climate-related risks for price stability. As can be seen from Figure \ref{fig:DevMaps}, both average mean temperatures in the reference period (1950-80) as well as average deviations in the last two decades varied considerably across the euro area. The average deviation from the historical mean ranges from $0.5^{\circ}C$  to $3^{\circ}C$, with the highest deviations in the southern and central regions of the euro area. Particularly in the larger (southern) countries such as Spain and Italy deviations from the historical mean vary markedly across regions. Moreover, the deviation from the historical mean is not uniform throughout the year. On average, deviations are highest in spring and lowest in autumn. Table \ref{country_climate_deviations} in the appendix shows the average yearly deviation from the historical mean for different euro area countries and weather variables. 

\bigskip

\FloatBarrier

\section{Weather Impact on Sectoral Inflation: Local Projections in a Low-Dimensional Context \label{LP}}

To set up a benchmark, we estimate the impact of different weather variables on prices in each sector separately using euro area averages for all variables considered. The effects are measured by impulse response functions (IRFs) derived from local projections. In the simplest (linear) form, we estimate the following model for each forecast horizon $h=0,1,2, ..., 24$,
\begin{equation}
y_{s,t+h}=\alpha_{0}^{h}+ \sum_{k=1}^pA_{k}^{h}y_{s,t-k}+\sum_{i=0}^rB_{i}^{h}x_{t-i}+ \sum_{j=1}^lC_{j}^{h}z_{s,t-j}+e_{t+h},
\label{Equ_LP}
\end{equation}
where $y_{s,t}$ is a vector of endogenous variables including year-on-year inflation rates of consumer and producer prices of sector $s$, (euro area) commodity prices and industrial production.\footnote{Appendix A provides a detailed description of the variables used and their respective transformations. In the subsector analysis the level of disaggregation might differ for the different variables included in $y_{s,t}$.} This means that we implicitly assume that there is no contemporaneous relationship between the elements of $y_{s,t}$. We consider up to $p$ lags of the endogenous variables, where the lag length $p$ is determined using the Akaike information criterion (AIC) and is therefore sector-specific. $x_{t}$ is the vector of weather shocks that may contemporaneously impact prices, and $z_{s,t}$ is a vector of control variables which includes weather variables other than the one(s) included in $x_{t}$\footnote{Similarly to \citet{KotzKuikLisNickel2023}, we use several combinations of different weather shocks and control for all weather variables.} and possibly relevant commodity prices (e.g. oil prices) so that $z_{s,t}$ is also sector-specific. The lag length $l$ is again determined by the AIC and may therefore depend on the sector being studied. $e_{t+h}$ are the innovations. The impulse response function of the endogenous variable at horizon $h$ to a change in weather is computed as, 
\begin{equation} 
\frac{\partial y_{s,t+h}}{\partial x_{t}} = B_{0}^{h}.
\end{equation}

For country-specific analysis we slightly modify (\ref{Equ_LP}) by replacing the variables in $y_{t,s}$ and $x_{t,s}$ by their respective country counterparts. Depending on the country of interest, commodity prices are considered to be exogenous if the country is sufficiently small. Given the results of \citet{CiccarelliKuikHernández2023} (see also \citealp{AhmadiCasoliManeraValenti2022}), we do not consider a panel setting, as countries even within the euro area seem to be heterogeneous in terms of their exposition and reaction to climate shocks; see e.g. \citet{KotzKuikLisNickel2023} and \citet{FacciaParkerStracca2021} for the impact of climate shocks in a panel framework.

\subsection{The definition of a weather shock}

A shock is defined as the deviation from the historical mean, where the latter is computed as the monthly average of the period from 1950 to 1980 and is depicted in Table \ref{table_historical_mean}. 

\begin{table}[bht!]
\caption{Historical mean (1950-1980) for average daytime temperature (in $^{\circ}C$), precipitation (in $mm/m^{2}$), solar radiation (in $W/m^{2}$) and wind speed (in $m/s$) for the euro area.}
\begin{center}
\scalebox{0.9}{\begin{tabular}{lcccccccccccc} 
\toprule
 \multicolumn{1}{c}{} & Jan & Feb & Mar & Apr & May & Jun & Jul & Aug & Sep & Oct & Nov & Dec \\ [0.5ex] 
 \midrule
 Temperature & -0.6 & 0.2 & 3.2 & 7.0 & 11.7 & 15.9 & 18.1 & 17.5 & 14.0 & 9.3 & 4.4 & 1.0 \\ 
 Precipitation & 1.9 & 1.9 & 1.7 & 1.7 & 1.8 & 2.0 & 1.8 & 1.9 & 2.0 & 2.1 & 2.4 & 2.2  \\
 Solar radiation & 49 & 78 & 127 & 180 & 230 & 246 & 246 & 214 & 156 & 99 & 57 & 42  \\
 Wind speed & 0.08 & 0.08 & 0.1 & 0.09 & 0.09 & 0.09 & 0.08 & 0.08 & 0.07 & 0.09 & 0.1 & 0.1  \\
 \bottomrule
\end{tabular}
 }
 \label{table_historical_mean}
 \end{center}
 \end{table}
\FloatBarrier

As shocks we only consider values that exceed a certain threshold, namely the average deviation in the period 2001-2021 from the historical mean. Table \ref{country_climate_deviations} in the Appendix shows these threshold values. For instance, for the euro area, we study the impact of deviations from historical mean temperatures exceeding 1.3$^{\circ}$ Celsius. Furthermore, empirical evidence highlights the importance of non-linearities in the impact of climate shocks. Hence, we consider several extensions to the baseline model. First, we study seasonal effects by isolating the impact of a weather shock on consumer prices for the four seasons of the year (spring (March-May), summer (June-August), autumn (September-November) and winter (December-February)). Second, we analyze whether consumer prices react differently to positive and to negative shocks. Third, we study extreme events by considering weather shocks that exceed a certain size in magnitude. Finally, different combinations of these non-linearities are explored. 

\subsection{Empirical results: strongest impact on food and energy}
\begin{figure}[htb]
    \centering 
\begin{subfigure}{0.49\textwidth}
  \includegraphics[width=\linewidth, height=5cm]{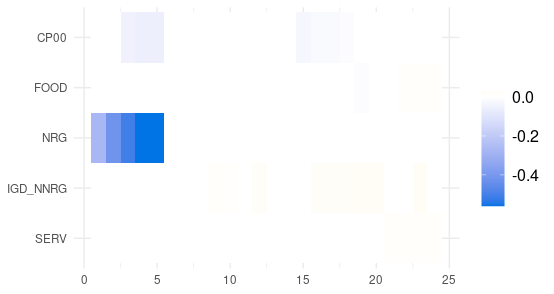}
  \caption{Temperature}
  \label{fig:1}
\end{subfigure}\hfil 
\begin{subfigure}{0.49\textwidth}
  \includegraphics[width=\linewidth, height=5cm]{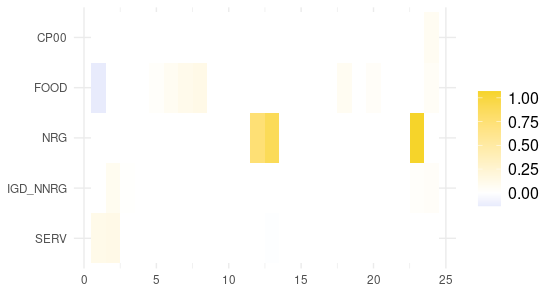}
  \caption{Precipitation}
  \label{fig:2}
\end{subfigure}\hfil 
\medskip

\begin{subfigure}{0.49\textwidth}
  \includegraphics[width=\linewidth, height=5cm]{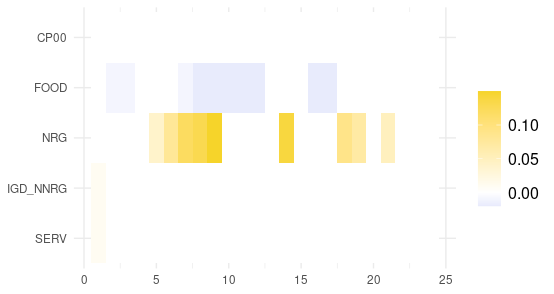}
  \caption{Minimum Temperature}
  \label{fig:3} 
\end{subfigure}\hfil
\begin{subfigure}{0.49\textwidth}
  \includegraphics[width=\linewidth, height=5cm]{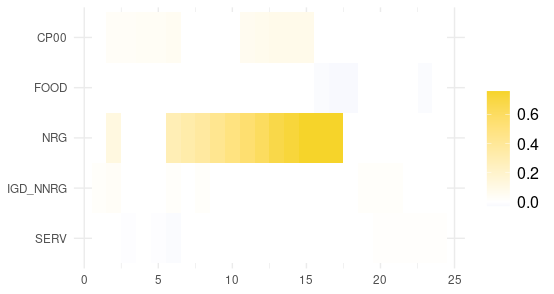}
  \caption{Maximum Temperature}
  \label{fig:4}
\end{subfigure}\hfil 
\medskip

\begin{subfigure}{0.49\textwidth}
  \includegraphics[width=\linewidth, height=5cm]{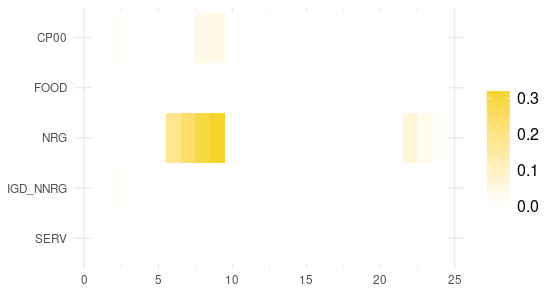}
  \caption{Wind Speed}
  \label{fig:5}
\end{subfigure}\hfil 
\begin{subfigure}{0.49\textwidth}
  \includegraphics[width=\linewidth, height=5cm]{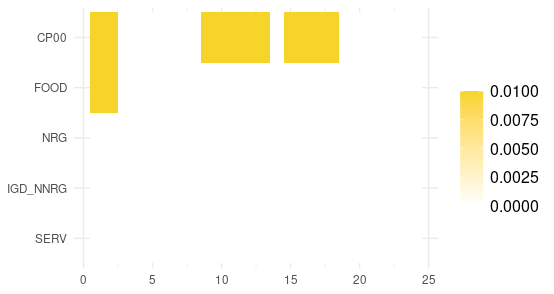}
  \caption{Solar Radiation}
  \label{fig:6}
\end{subfigure}\hfil 
\caption{Impact of a deviation from the historical mean on special aggregates and headline inflation (CP00) in percentage points for the euro area in the 24 months after the weather shock.}
\label{fig:EA_SA}
\end{figure}

 We start by calculating the average impact of a weather shock on HICP inflation in 89 subsectors for the euro area. We compare different specifications as far as weather variables and oil prices are concerned, but always include producer prices (except for the services sector), industrial production and commodity prices. The impact on inflation rates varies considerably across subsectors with the effect on aggregate inflation measures, such as food, energy, services and non-energy industrial goods being muted. Figure \ref{fig:EA_SA} shows the impact of a deviation from the historical mean for different weather variables for the euro are in percentage points up to 24 months after a shock. For all weather variables except minimum temperature, the impact on headline inflation (CP00) is statistically significant, but small. Only average temperature shocks have disinflationary impacts, while higher wind speeds, solar radiation or precipitation tend to raise inflation in the months after the shock. Energy consumer prices react more strongly than food prices to weather shocks, while core inflation (services and non-energy industrial goods) is barely affected by weather anomalies.

\FloatBarrier

\begin{figure}[bht!]                 
  \centering
  \includegraphics[width=16cm, height=8cm]{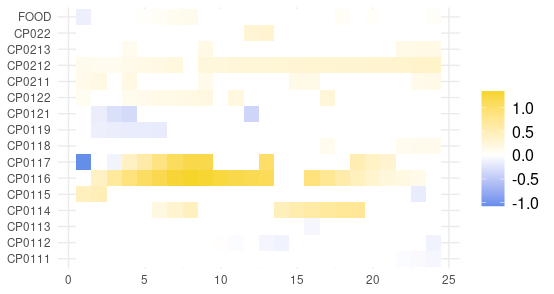}
  \caption{Average impact of a precipitation shock on euro area HICP inflation rates in the food sector measured in percentage points (90\% confidence interval). The x-axis shows the horizon in months and the y-axis the different food (sub-)sectors. Controls: Temperature, wind speed, oil prices.}
  \label{fig:FOOD_Precipitation}
\end{figure}

While the impact on inflation rates is small if we focus on the four special aggregates, certain subsectors are considerably affected. The effects are heterogeneous across sectors and in the aggregate often offset each other or become negligible. Similarly, the time lag with which shocks impact consumer prices varies markedly across subsectors. Most effects are observed after 6 to 15 months, but in some subsectors such as non-alcoholic beverages or fuels, we also find short-term effects. Figure \ref{fig:FOOD_Precipitation} illustrates some of the results by showing the effects of a precipitation shock on the inflation rates of all sectors related to food including alcohol and tobacco. The strongest impact is observed in category CP0116 - fruits, but we also find inflationary effects on the inflation rate of wine (CP0212) or milk products (CP0114).  

\FloatBarrier

In Appendix \ref{AppB} we present more detailed figures for different combinations of sectors, weather shocks and control variables. The results show that the effects depend heavily on the choice of the control variables that are included in the analysis. For the food sector, for example, we find that the effects on food inflation can be positive as well as negative depending on the choice of control variables. Another interesting observation can be made when comparing the effects of temperature deviations. For energy, for instance, deviations in minimum or maximum temperatures have inflationary effects, whereas shocks to average temperature deviations have a negative impact on energy inflation.  

\subsubsection{Negative shocks}

\begin{figure}[htb]
    \centering 
\begin{subfigure}{0.49\textwidth}
  \includegraphics[width=\linewidth, height=5cm]{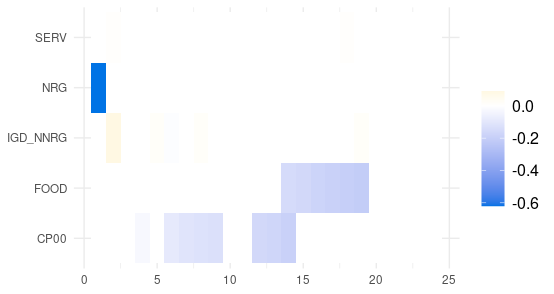}
  \caption{Temperature}
  \label{fig:EA_SA_Temp_neg}
\end{subfigure}\hfil 
\begin{subfigure}{0.49\textwidth}
  \includegraphics[width=\linewidth, height=5cm]{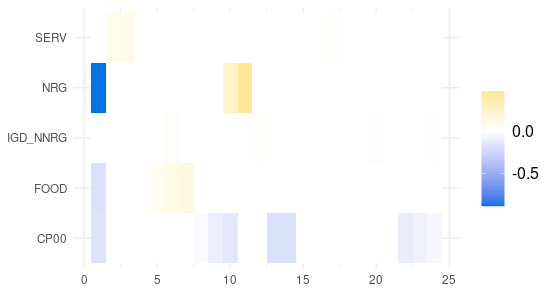}
  \caption{Precipitation}
  \label{fig:EA_SA_Prec_neg}
\end{subfigure}\hfil 
\medskip
\caption{Impact of a negative deviation from the historical mean on special aggregates and headline inflation (CP00) in percentage points for the euro area.}
\label{fig:EA_SA_neg}
\end{figure}

So far, we showed the results for positive deviations from the mean such as unusually high temperatures or precipitation. Figure \ref{fig:EA_SA_neg} shows the effects of a negative weather shock for average daytime temperatures and for precipitation. The impact on headline inflation is small and negative. A comparison to figure \ref{fig:EA_SA} shows that the effects differ considerably from the ones we observe when the shock is positive. This difference is not simply in terms of the sign, but also in terms of sectors, timing and magnitude of the effect. This confirms that weather anomalies impact consumer prices in a non-linear fashion.

\subsubsection{Seasonal effects}

\begin{figure}[bht!]                 
  \centering
  \includegraphics[width=15cm]{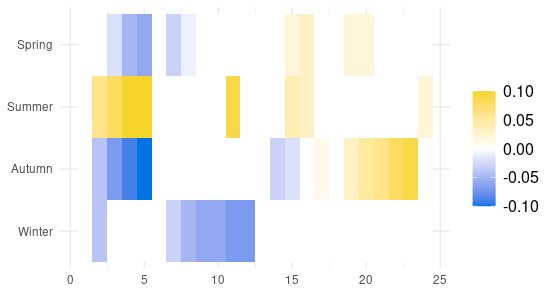}
  \caption{Seasonal effects of a temperature shock on food inflation.}
  \label{fig:food_season_temp}
\end{figure}

Figures \ref{fig:EA_SA} and \ref{fig:FOOD_Precipitation} show the average impact throughout the year. However, the timing of a weather shock strongly influences its impact on prices, as can be seen in Figure \ref{fig:food_season_temp}. A temperature shock in summer lifts consumer price inflation for food in the months following the shock, while a shock in winter has a negative impact. In spring and autumn the impact changes from negative in the short run to positive in the medium run. If we compare these findings to Figure \ref{fig:1}, we see that the overall impact of a temperature shock on food prices being insignificant is a result of countervailing impacts across the different seasons.

\FloatBarrier

\subsubsection{Geographical heterogeneity}

\begin{figure}[bht!]                 
  \centering
  \includegraphics[width=16cm, height=7cm]{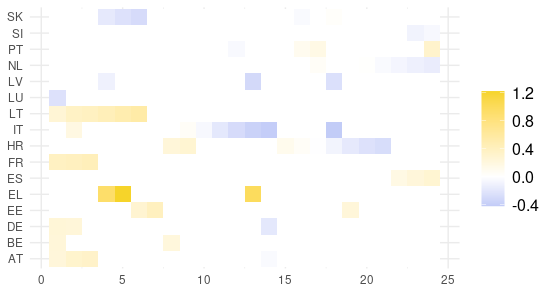}
  \caption{Impact of a temperature shock in summer on energy inflation in different euro area countries.}
  \label{fig:countries_NRG_summer}
\end{figure}

Focusing on euro area aggregates conceals a large heterogeneity across countries. Not only do countries differ in terms of their exposure to weather shocks and climate change, also differences in economic structures may amplify or dampen the impacts on consumer prices. Figure \ref{fig:countries_NRG_summer} depicts the reaction of energy inflation across various euro area member states to a temperature shock in summer. In some countries this shock is inflationary, in others disinflationary. Similarly the timing and the size of the impact vary markedly. Differences in the energy mix or price setting mechanisms in the energy markets (incl.  taxation or subsidies by national governments) might be responsible for the large heterogeneity together with climatic differences. 
\FloatBarrier



The preceding subsections underscored the significant heterogeneity in the impact of weather on inflation rates across subsectors, seasons, and countries. However, when aggregating over sectors, time, or regions, valuable information about the economic consequences of extreme weather events is lost. In the next section, we propose a method to extract the relevant information from the surface data for our analysis.

\section{Relating high and infinite-dimensional data: Associated factors \label{Associated_factors}}
Understanding and modelling relationships between changes in price indices and changes in climate variables presents significant challenges. With price indices consisting of 89 different sectors and climate data treated as time series of surfaces, which implies infinite dimensionality, it can be difficult to identify and quantify patterns or relationships between these data sets.
Complications are further amplified by the fact that not all indices may be affected, only a subset of them. Furthermore, not all information carried by climate variables is relevant for price developments. Thus, how can we identify the ``relevant'' part? To understand and model the relationships between these two data sets, we first identify the parts in both data sets that relate to each other. In what follows we will refer to these as \textcolor{black}{associated factors}. Once identified, appropriate methodological techniques have to be developed to extract and estimate the \textcolor{black}{associated factors} from both data sets, and provide a subsequent analysis. For the latter, we propose a generalized impulse-response analysis that enables us to exploit the geographical aspect of the climate data.  

To offer an intuitive explanation of the methodology, we start our discussion with an informal description of the problem. We have one finite-dimensional variable $Y$, representing
changes in fixed and finite number of consumer prices, and one infinite-dimensional variable $X$, representing changes in a climate variable. We are interested decomposing $Y$ and $X$ such that
\begin{equation}\label{eq:CommonFactors}
    Y = \Tilde{Y} + Y^\perp \text{ and }  X = \Tilde{X} + X^\perp,    
\end{equation}
where $\Tilde{Y}$ and $\Tilde{X}$ represent the parts referred to as \textcolor{black}{associated factors}, encapsulating the essence of the relationship between X and Y, whereas $Y^\perp$ and $X^\perp$ are irrelevant for describing this relationship. These associated factors can then in a second step be used for a (linear) structural representation, which may be of interest for a subsequent analysis. This representation is given by
\begin{equation}\label{eq:StructuralModel}
     \Tilde{Y} =\mathbf{\Pi}\left(\Tilde{{X}}\right) + {e},
\end{equation}
where $e$ is an idiosyncratic error term. In Section \ref{sec:FIRA}, we show how representation \eqref{eq:StructuralModel} is used to construct functional impulse-responce analysis (FIRA). The term ``functional'' refers here to the nature of climate variables that are treated as surfaces. FIRA allows us to control for the magnitude of the shock as well as for its geographical location and size of the area. 

\subsection{Identification of Associated Factors}

In the case when both $X$ and $Y$ are finite dimensional, the solution to the problem \eqref{eq:CommonFactors}-\eqref{eq:StructuralModel}  is well-known and can be derived through canonical correlation analysis (CCA). CCA is a widely-used method in conventional (finite-dimensional) settings for analyzing linear associations between two datasets, as initially formulated in a seminal work by Hotelling (1936). However, extending this methodology to our problem is challenging, primarily due to the infinitely dimensional nature of the variables representing climate conditions. To explain the complexities introduced by this characteristic and to provide insights into our solution, we begin by describing the case where both sets of variables are finite-dimensional.

\subsubsection*{The finite-dimensional case}

Consider two random vectors $Y$ and $X$ in $\mathbb{R}^p$ and $\mathbb{R}^q$, respectively. The problem of finding the associated factors as in \eqref{eq:CommonFactors} can be formulated as finding the pairs of directions  $a_i\in\mathbb{R}^p$ and $b_i\in\mathbb{R}^q$, for $i=1,...,K\leq \min(p,q)$, across which $Y$ and $X$ have nonzero covariances. That is,
\begin{equation} 
    Cov(\langle a_i,Y \rangle \langle b_i,X\rangle)\neq0,
\end{equation}
where $\langle .,. \rangle$ denotes the inner product (or the scalar product) on $\mathbb{R}^p$ or $\mathbb{R}^q$ and can be interpreted as a projection of $Y$ into $a_i$ (or $X$ into $b_i$)\footnote{Here, with the small abuse of notation, we use the same symbol for the inner product on $\mathbb{R}^p$ and $\mathbb{R}^q$.}. Furthermore, pairs $(a_i,b_i)$ are selected in a meaningful way, such that, the correlation between $\langle a_i,Y \rangle$ and $\langle b_i,X\rangle$ is maximized and each pair $(a_i,b_i)$ is orthogonal to $(a_j,b_j)$ for $i\neq j$. Here  orthogonality is defined as $Cov(\langle a_i,Y \rangle)\langle b_j,X\rangle)=Cov(\langle a_i,Y \rangle)\langle a_j,Y\rangle) =$ $Cov(\langle b_i,X \rangle)\langle b_j,X\rangle)=Cov(\langle b_i,X \rangle)\langle a_j,Y\rangle)=0$. From this we obtain the structural relationship between $Y$ and $X$ as,
\begin{equation}\label{eq:StructuralModelFinite}
     \Tilde{\mathbf{Y}} =\mathbf{\Pi}\Tilde{\mathbf{X}} + \mathbf{e},
\end{equation}
where $\Tilde{\mathbf{Y}}=[\langle a_1,Y \rangle,...,\langle a_K,Y \rangle]'$, $\Tilde{\mathbf{X}}=[\langle b_1,X \rangle,...,\langle b_K,X \rangle]'$, $\mathbf{\Pi}=\mathrm{diag}(\rho_1,...,\rho_K)$ and $\rho_k$ denotes the  correlation $\text{corr}(\langle a_k,Y \rangle \langle b_k,X\rangle)$. The error term $\mathbf{e}$ is a $(K\times1)$ vector of errors that is uncorrelated with $\Tilde{\mathbf{X}}$. 

In other words we are searching for solutions $\{\rho_k,a_k,b_k\}_{k=1}^K$ that solve the maximization problem,
\begin{equation}\label{eq:maxcorr}
   \rho_k= \max_{\tilde{a}_k\in\mathbb{R}^p\,\text{, }  \tilde{b}_k\in\mathbb{R}^q} \text{corr}(\langle \tilde{a}_k,Y \rangle \langle \tilde{b}_k,X\rangle),
\end{equation}
subject to the orthogonality condition on $\{a_k,b_k\}_{k=1}^K$.\footnote{We first maximize among all choices in $\mathbb{R}^p$ and $\mathbb{R}^q$ to find $a_1$ and $b_1$. To find $a_2$ and $b_2$ subject to the orthogonality restriction implies maximization among all choices in $\mathbb{R}^p$ and $\mathbb{R}^q$ orthogonal to $a_1$ and $b_1$. This results in $\rho_2\leq\rho_1$ as $\rho_2$ is the maximum over a smaller subspace. By continuing this process we can find $K\leq \min(p,q)$ couples.} The solution to these maximization problems is found through the singular-value decomposition (SVD) of the correlation operator (see, for instance,  \citealp{JohnsonWichern2009}) 
\begin{equation} \label{eq:CorrOper}
   \mathcal{M} = C_X^{-1/2} C_{XY} C_Y^{-1/2},
\end{equation}
where $C_{X}=Var[X]$, $C_Y=Var[Y]$ and $C_{XY}=Cov[X,Y]$. The SVD yields a set of singular value triplets $\{\rho_k,f_k,g_k\}_{k=1}^K$, 
where , $C_X^{-1/2} C_{XY} C_Y^{-1/2}(f_k)=\rho_k g_k$. Furthermore, $\rho_k$ is the solution to the maximal correlation problem in \eqref{eq:maxcorr} and the corresponding directions are given as $a_k= C_Y^{-1/2}f_k$ and $b_k=C_X^{-1/2}g_k$, for $k=1,...,K$. Once all $K$ couples, $\{a_k,b_k\}_{k=1}^K$, are found we can rewrite the original two sets of variables as, 
\begin{equation*}
    Y = \sum_{k=1}^K \langle a_k,Y \rangle a_k + Y^\perp \text{ and }  X = \sum_{k=1}^K \langle b_k,X \rangle b_k + X^\perp,    
\end{equation*}
where $Y^\perp$ is the part of $Y$ that is uncorrelated with $X$ and 
$X^\perp$ is the part of $X$ that is uncorrelated with $Y$. Hence, $\sum_{i=1}^K \langle a_i,Y \rangle a_i$ and $\sum_{i=1}^K \langle b_i,X \rangle b_i$ are identified, respectively, as the \textcolor{black}{associated factors}, $\Tilde{Y}$ and $\Tilde{X}$, described in \eqref{eq:CommonFactors}. Furthermore, they can be written in the structural form \eqref{eq:StructuralModelFinite}.

\subsubsection*{Mix of Finite and Infinite Dimensional Variables}

In the context of our application $X$ represents a climate variable (e.g. average daytime temperature) which we treat as a surface over the geographical domain of the euro area. 
From a technical perspective, these surfaces are functions within a well-defined functional space, implying that, in general, they are infinitely-dimensional objects.
The primary challenge in extending finite-dimensional Canonical Correlation Analysis (CCA) to our case arises from the fact that the counterpart of the covariance $C_X$ in \eqref{eq:CorrOper} is not invertible, constituting what is known as an ill-posed inverse problem.
Therefore, the first crucial point is to verify the existence of \textcolor{black}{associated factors}. Second, in empirical applications, some form of regularization of the problem is required to derive an appropriate estimator. 

To prove the existence of associated factors, we adapt the results available in the functional data analysis on CCA. \cite{HE2002} considers an extension of CCA to the case when data samples consist of pairs of square integrable stochastic processes. The authors derive conditions on the dependence between $Y$ and $X$ (see \citealp[Condition 4.5]{HE2002}) to ensure the existence of the correlation operator in \eqref{eq:CorrOper}. An alternative approach proposed in \cite{EubankHsing2008} transforms the functional CCA problem into finding so-called canonical functions in the reproducing kernel Hilbert space generated by the corresponding covariance functions. \cite{Kupresanin2010} provides a theoretical comparison of both approaches. We use tools from both papers to derive a sufficient condition under which \textcolor{black}{associated factors} exist and are well defined.

When $X$, $Y$ and the associated factors are infinite-dimensional, the "regularization" of the problem cannot be avoided. This point is also stressed in \cite{Leurgans1993}. However in our case, the regularization can circumvented, since $Y$ is finite dimensional and therefore the associated factors are finite dimensional by construction. 
This leads to the fact that both operators $\mathcal{M}$ and $C_{YX}$ are finite dimensional and SVD of both produces right and left singular vectors that span the same subspaces (provided that $\mathcal{M}$ exists and is well defined). 
That is,  $\{\alpha_i,\beta_i\}_{i=1}^K$, extracted from SVD of $C_{XY}$ generate subspaces  $\Omega_\alpha=span\{\alpha_1,...,\alpha_K\}$ and  $\Omega_\beta=span\{\beta_1,...,\beta_K\}$. These subspaces are identical to the corresponding subspaces produced by the solutions to the maximum correlation problem,  $\Omega_a=span\{a_1,...,a_K\}$ and  $\Omega_b=span\{b_1,...,b_K\}$. This in turn implies that projections of the original $Y$ and $X$ into $\Omega_\alpha=span\{\alpha_1,...,\alpha_K\}$ and  $\Omega_\beta=span\{\beta_1,...,\beta_K\}$ allow us to extract common factors  as $\Tilde{Y}=\sum_{k=1}^K \langle \alpha_k,Y \rangle \alpha_k$ and $\Tilde{X}=\sum_{k=1}^K \langle \beta_k,X \rangle_H \beta_k$. As a result we obtain a simple and straightforward approach to identify and estimate associated factors based on the SVD of $C_{YX}$ rather than the SVD of $\mathcal{M}$ which involves inverses of $C_{X}$ and the consequent estimator regularization. Furthermore, to construct a structural model as in \eqref{eq:StructuralModel} we can first extract projections $\Tilde{Y}_K$ and $\Tilde{X}_K$ as described above and then perform the conventional CCA on these elements as they are finite-dimensional by construction. This result is formalized below.

To define and derive the  \textcolor{black}{associated factors} additional notation is required. We work with two random variables, $Y$ and $X$, that take values in $\mathbb{R}^p$ and in $H\equiv L^2(\Delta,\langle ,\rangle_H)$, respectively. Here $H$ is the space of square-integrable real-valued functions, which is equipped with the standard inner product $\langle x,y\rangle_H=\int_{\Delta}x(u)y(u)\mathrm{d}u$, and $\Delta \subset \mathbb{R}^2$ is the domain of functions on $H$ that is associated with the geographical restrictions of the data (i.e., geographical area of euro zone). Since \textcolor{black}{associated factors} are determined solely by the covariance structure, we define the following covariance operators for $x \in H$ and $y \in \mathbb{R}^p$,
\begin{eqnarray*}
C_{X}(x) =&E\left[ \langle X,x \rangle_H X\right]; &\, C_{XY}(x)= E\left[ \langle X,x \rangle_H Y\right];\\
C_{YX}(y)=& E\left[ \langle Y,y \rangle X\right]; & \,
C_{Y}(y)= E\left[ \langle Y,y \rangle Y\right]. 
\end{eqnarray*}
Furthermore, $C_{X}:H\rightarrow H$, $C_{Y}:\mathbb{R}^p\rightarrow \mathbb{R}^p$, $C_{XY}:H\rightarrow \mathbb{R}^p$ and $C_{YX}:\mathbb{R}^p\rightarrow H$ and each of them admits the spectral decomposition:
\begin{eqnarray*}
C_{X}(x) &=& \sum_{i=1}^\infty \lambda_i \langle x,\phi_i \rangle_H \phi_i;\\
C_{Y}(y) &=& \sum_{j=1}^p \gamma_j \langle x,\psi_j \rangle \psi_j;\\
C_{YX}(y) &=& \sum_{k=1}^K r_k \langle y,\alpha_k \rangle \beta_k.
\end{eqnarray*}
The associated factors $\Tilde{Y}=\sum_{k=1}^K \langle {a}_k,Y \rangle {a}_k$ and $\Tilde{X}=\sum_{i=1}^K \langle {b}_i,X \rangle {b}_i$ are defined through, $(\rho_k,a_k,b_k)_{k=1}^K$, the solution to the maximization problem 
\begin{equation} \label{eq:FCCA}
   \rho_k= \sup_{a\in\mathbb{R}^p\,\text{, }  b\in H} \text{Cov}(\langle a,Y \rangle \langle b,X\rangle_H), \text{ for } k=1,...,K
\end{equation}
subject to $Var(\langle a,Y \rangle)=1$ and $Var(\langle b,X\rangle_H)=1$ and orthogonality condition between couples $(a_i,b_i)$ is orthogonal to $(a_j,b_j)$ .  Finally, we define the covariance operators for $\Tilde{Y}$ and $\Tilde{X}$ as $\Tilde{C}_{X}(x) =E\left[ \langle \Tilde{X},x \rangle_H \Tilde{X}\right]$, $\Tilde{C}_{Y}(y)= E\left[ \langle Y,y \rangle Y\right]$ and $\Tilde{C}_{YX}(y)= E\left[ \langle \Tilde{Y},y \rangle \Tilde{X}\right]={C}_{YX}(y)$.

\begin{theorem} \label{thm:Identification}
    Under the regularity condition 
    \begin{equation}\label{eq:RegCond}
     \sum_{i=1}^\infty    \lambda_i^{-1} E[\langle X,\phi_i \rangle_H \langle Y,\psi_j \rangle]<\infty \text{ for all } j, 
    \end{equation}
    we have that 
    
    \begin{description}
      \item[(i)] the solution $\{\rho_k,a_k,b_k\}_{k=1}^K$ to maximization problem \eqref{eq:FCCA} exists and is expressed through SVD  $\{\rho_k,f_k,g_k,\}_{k=1}^K$ of the operator $C_X^{-1/2} C_{XY} C_Y^{-1/2}$ for $k=1,...,K$ as \[\{\rho_k,C_Y^{-1/2}(f_k),C_X^{-1/2}(g_k)\}_{k=1}^K,\]
      where $C_X^{-1/2} C_{XY} C_Y^{-1/2}(f_k)=\rho_k g_k$.
      \item[(ii)] $span\{a_1, ..., a_K\}=span\{\alpha_1,...,\alpha_K\}$ and  $span\{b_1, ..., b_K\}=span\{\beta_1,...,\beta_K\}$;
      \item[(iii)] the solution, $(\rho_k,a_k,b_k)_{k=1}^K$, can be equivalently obtained from SVD of the operator $C_{\Tilde{X}}^{-1/2} C_{\Tilde{X}\Tilde{Y}} C_{\Tilde{Y}}^{-1/2}$.
    \end{description}
\end{theorem}

Theorem \ref{thm:Identification} solves several important issues in identifying associated factors and the structural representation of their relationship. First, the existence of the associated factors, challenged by the ill-posed inverse problem on $C_X$, is obtained under the regulatory condition \eqref{eq:RegCond}. This condition ensures that the range of the operator $C_{YX}$ is in the range of the operator $C_X^{1/2}$, which in turn implies that the operator $C_X^{-1/2} C_{XY} C_Y^{-1/2}$ is well defined. Further technical discussion on the role of \eqref{eq:RegCond} is given in the remark below. Second, Theorem \ref{thm:Identification} shows in item (ii) that the associated factors, if they exist, can be obtained from projecting into  $\Omega_\alpha=span\{\alpha_1,...,\alpha_K\}$ and $span\{\beta_1,...,\beta_K\}$ (instead of $span\{a_1, ..., a_K\}$ and  $span\{b_1, ..., b_K\}$). Finally, the synthesis of items (ii) and (iii) offers a rationale for an alternative approach to acquiring the structural representation of the relationship. This involves projecting the original variables into $\Omega_\alpha$ and $\Omega_\beta$, thereby diminishing the dimensionality of the original variables and eliciting associated factors. Subsequently, employing CCA on these associated factors yields the desired structural representation. 

\subsection{Functional Impulse-Response Analysis} \label{sec:FIRA}
\begin{figure}
    \centering
    \includegraphics[width=\textwidth]{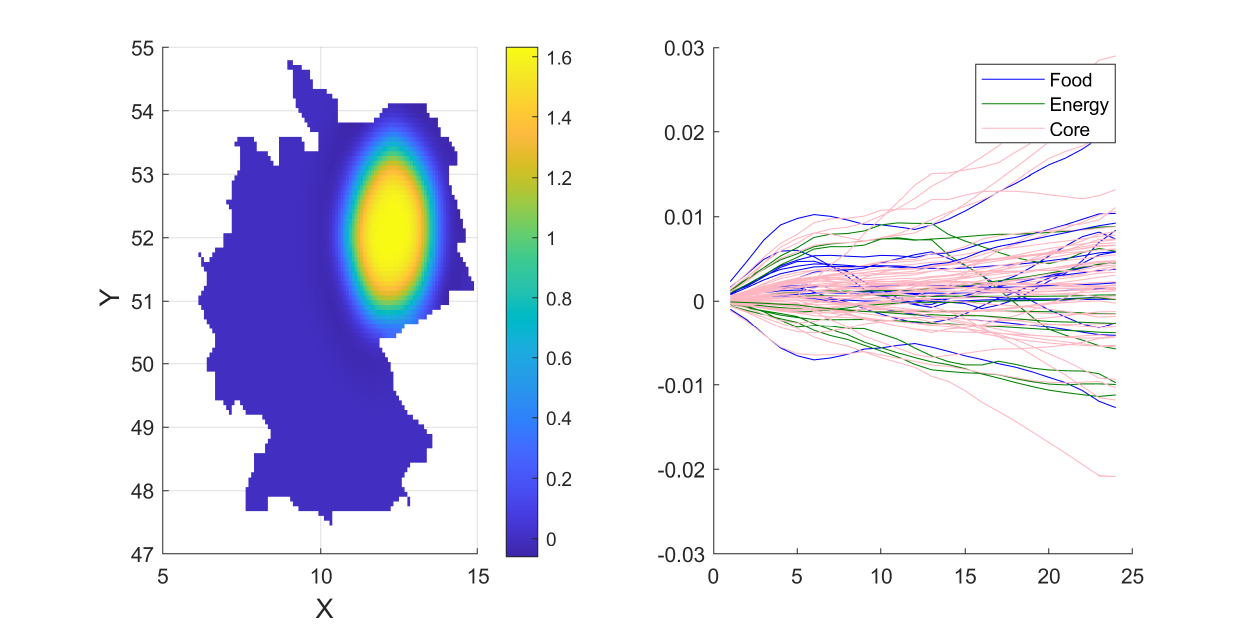}
    \caption{\footnotesize{Functional Impulse-Response Analysis to 1.5 $^\circ$C Temperature Shock in the North-East of Germany.}}
    \label{fig:ExampleFIRA}
\end{figure}

The structural representation in \eqref{eq:StructuralModelFinite} serves as the primary building block in the subsequent analysis. In this section, we utilize it to construct a functional impulse-response analysis (FIRA). To this end, we have discussed the identification of associated factors from one finite and one infinite-dimensional dataset. Now, we incorporate the time series feature of the data into our analysis. Let $Y_t\in \mathbb{R}^p$ denote a vector of price changes at time $t$, $X_t \in H$ is an exogenous climate shock at time $t$. We may be interested in including additional control variables in our analysis, such as lags of $Y_t$, lags of $X_t$ or other climate variable(s) as well as economic variables and their lags, which we denote as $Z_t \in H_Z\equiv L^2(\Delta_Z,\langle ,\rangle_{H})$.\footnote{An additional climate variable is not necessarily required to be defined over the same geographical area as $X_t$. However, it is essential that it belongs to the space of square-integrable functions with the standard inner product.} 
Let $V_t$ denote the collection of all relevant variables i.e.,
\begin{equation*}
    V_t=\left[X_t,X_{t-1},...,X_{t-q},Y_{t-1},...,Y_{t-s},Z_{t-1},Z_{t-l}\right].
\end{equation*}
In what follows we assume that all variables are weakly (covariance) stationary. To construct functional impulse-responses we extract associated factors between $Y_t$ and $V_{t-h}$ with $h=0,...,h_{max}$. That is, for each horizon $h$ we construct $\{\rho_k^{(h)},a_k^{(h)},b_k^{(h)}\}_{k=1}^{K_h}$ and 
\begin{equation*}
     \Tilde{\mathbf{Y}}_t =\mathbf{\Pi}_h\Tilde{\mathbf{V}}_{t-h} + \mathbf{e}_t,
\end{equation*}
where $\mathbf{\Pi}_h=diag(\rho_1^{(h)},...\rho_{K_h}^{(h)})$. The triplet $\{\mathbf{\Pi}_h,a_k^{(h)},b_k^{(h)}\}_{k=1}^{K_h}$ defines FIRA and can be seen as a generalization of the classical impulse-response analysis obtained via local projections. To be more precise, it allows us to operate with a generalized concept of a weather shock, where we control not only for the magnitude of the shock (e.g., $1.5 C^\circ$ deviation from the historical mean) but also for its geographical attributes such as location and coverage area. For instance, Figure \ref{fig:ExampleFIRA} plots the FIRA of a temperature shock on 80 sectors of the German economy. The temperature shock, plotted on the left panel, has the following attributes: magnitude - $1.5 C^\circ$; located Noth-East of Germany and covering an area with a radius of 150 km which corresponds to approx. 70700 square kilometers.


\section{Discussion \label{conclusion}}

The impact of climate change on consumer prices is a complex and multifaceted issue, as highlighted in this article. Its impact is heterogeneous across regions and sub-sectors, and is influenced by various weather variables with differing magnitudes, directions, and timing. Moreover, in the aggregate some of these effects offset each other or become negligible. This does not mean that the heterogeneity and complexity of these effects is not important for future price developments.

One crucial aspect to consider is the distinction between demand-driven and supply-driven effects. Certain weather events may lead to changes in consumer behavior and preferences, creating shifts in demand for specific goods and services. Simultaneously, other effects may disrupt the production and supply chains, resulting in fluctuations in the availability and cost of products. These supply-demand imbalances need to be carefully assessed and monitored as they may potentially create bottlenecks and lead to more widespread price pressures.

Additionally, the dynamic nature of relative prices in response to climate change should not be overlooked. As some sectors and regions are more susceptible to climate-related risks than others, relative prices may fluctuate. This may impact consumers differently across countries and across various categories of goods and services. Thereby it may also affect inflation heterogeneity in the euro area. A higher frequency of extreme weather events may also impact the volatility of inflation rates making it even harder to predict future inflation developments. 

Given the limitations outlined in this study, such as the focus on the euro area and the relatively short time period considered for price data, it is essential to recognize that the full extent of climate change's impact on consumer prices may not be fully captured. Other research has indicated that the impact in developing countries might be more pronounced, underscoring the need for a comprehensive global perspective on this issue. Furthermore, with the limitations of historical data, the analysis might not fully capture long-term trends or account for the potential amplification of climate-related effects over time.

In conclusion, it is crucial to harness all available information and consider climate change in all its dimensions when evaluating its impact on consumer prices. This will enable us to accurately identify potential areas where intervention may be necessary or beneficial in a timely manner. As climate change continues to exert its influence on our world, understanding its intricate relationship with consumer prices remains a critical aspect of (monetary) policy-making.

\newpage

\bibliographystyle{chicago}
\bibliography{ClimateReferences}

\newpage

\appendix
\section{Appendix: Data preprocessing}

\subsection*{Economic Data}
All data are used in year-on-year growth growths unless explicitly stated otherwise. Subsectors which do not have observations dating back to 2001 are removed for the analysis in Section \ref{Associated_factors}, but are included in sector by sector analysis of Section \ref{LP}. 

\subsection*{Climate Data}
Several data preprocessing steps are necessary before the E-OBS data can be used in our analysis. As the station density varies across time, we exclude all stations from the dataset which have missing observations. 

\section{Appendix: Figures and Tables \label{AppB}}
\renewcommand{\thefigure}{A.\arabic{figure}}%
\renewcommand{\thetable}{A.\arabic{equation}}%
\setcounter{figure}{0}
\setcounter{equation}{1}

\begin{landscape}
\captionsetup[subfigure]{aboveskip=1pt}
\begin{figure*}[th]
	\centering
	\begin{subfigure}[t]{0.8\textwidth}
		\includegraphics[width=12.5cm, height=15cm]{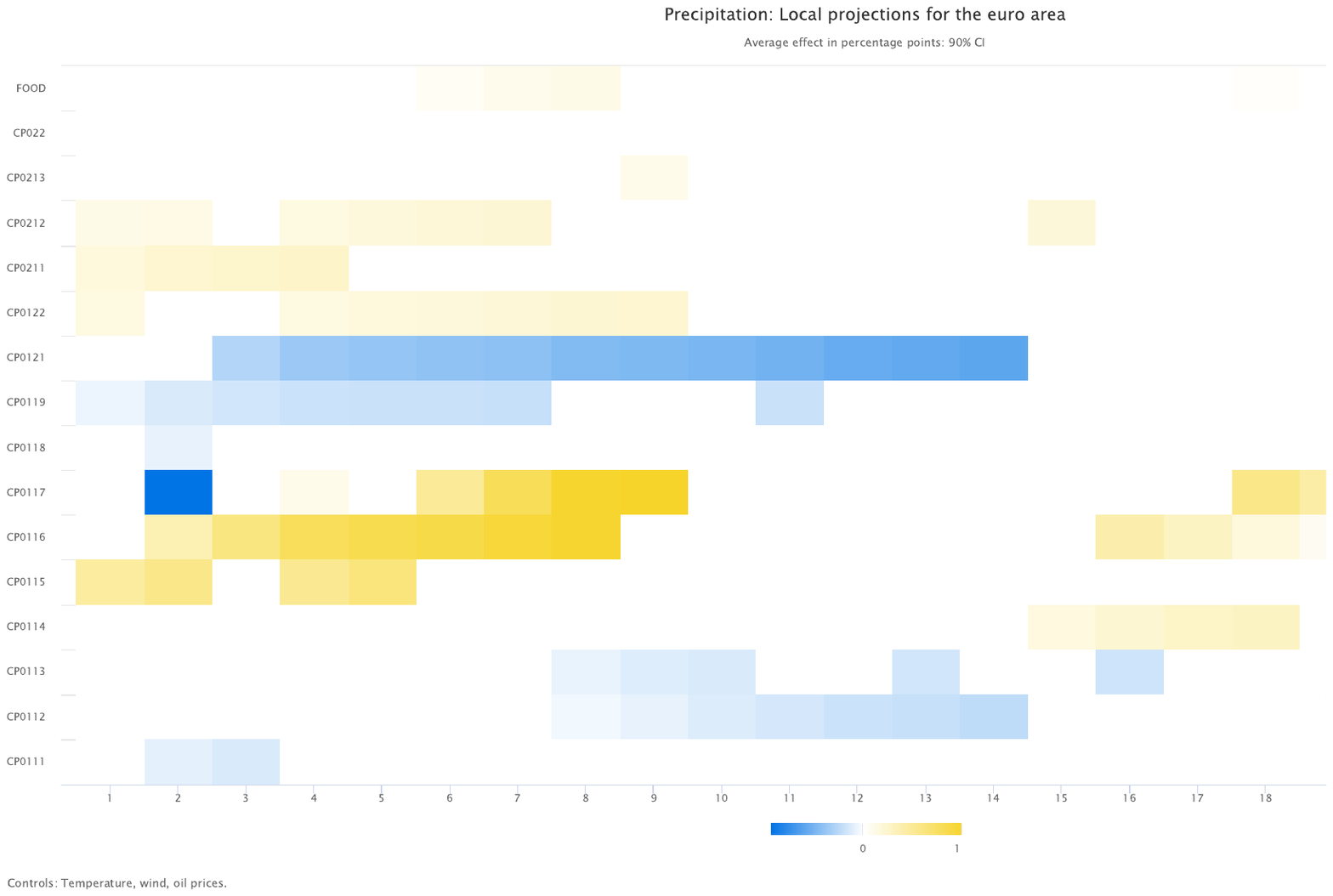}
	\end{subfigure}%
	\begin{subfigure}[t]{0.8\textwidth}
		\centering
		\includegraphics[width=12.5cm, height=15cm]{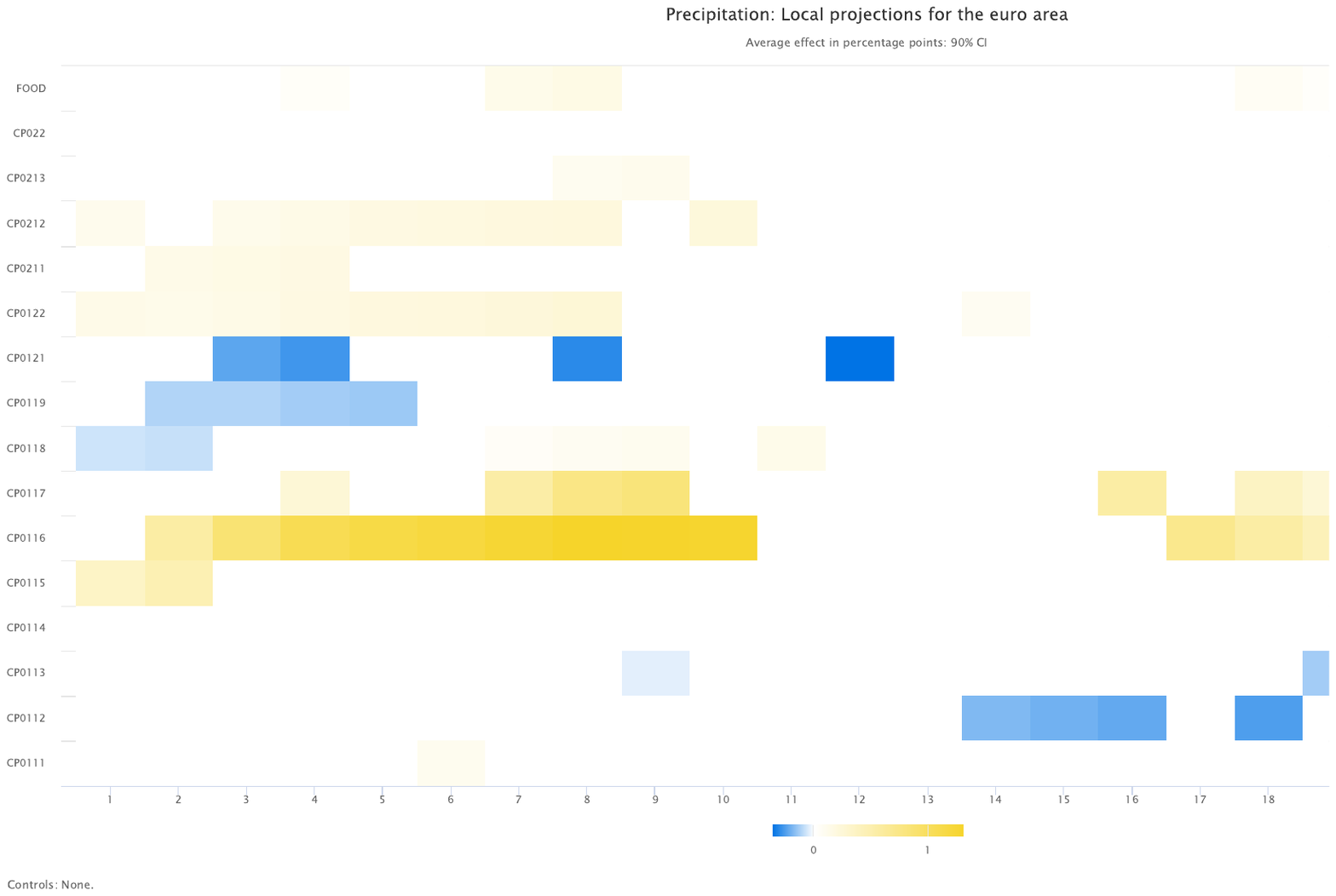}
	\end{subfigure}%
\vspace{-0.1cm}	
	\caption{Precipitation: Local projections for the Euro Area (Food)}
		 \label{fig:FigA1}
\end{figure*}	
\end{landscape}

\begin{landscape}
\captionsetup[subfigure]{aboveskip=1pt}
\begin{figure*}[th]
	\centering
	\begin{subfigure}[t]{0.8\textwidth}
		\centering
		\includegraphics[width=13cm, height=15cm]{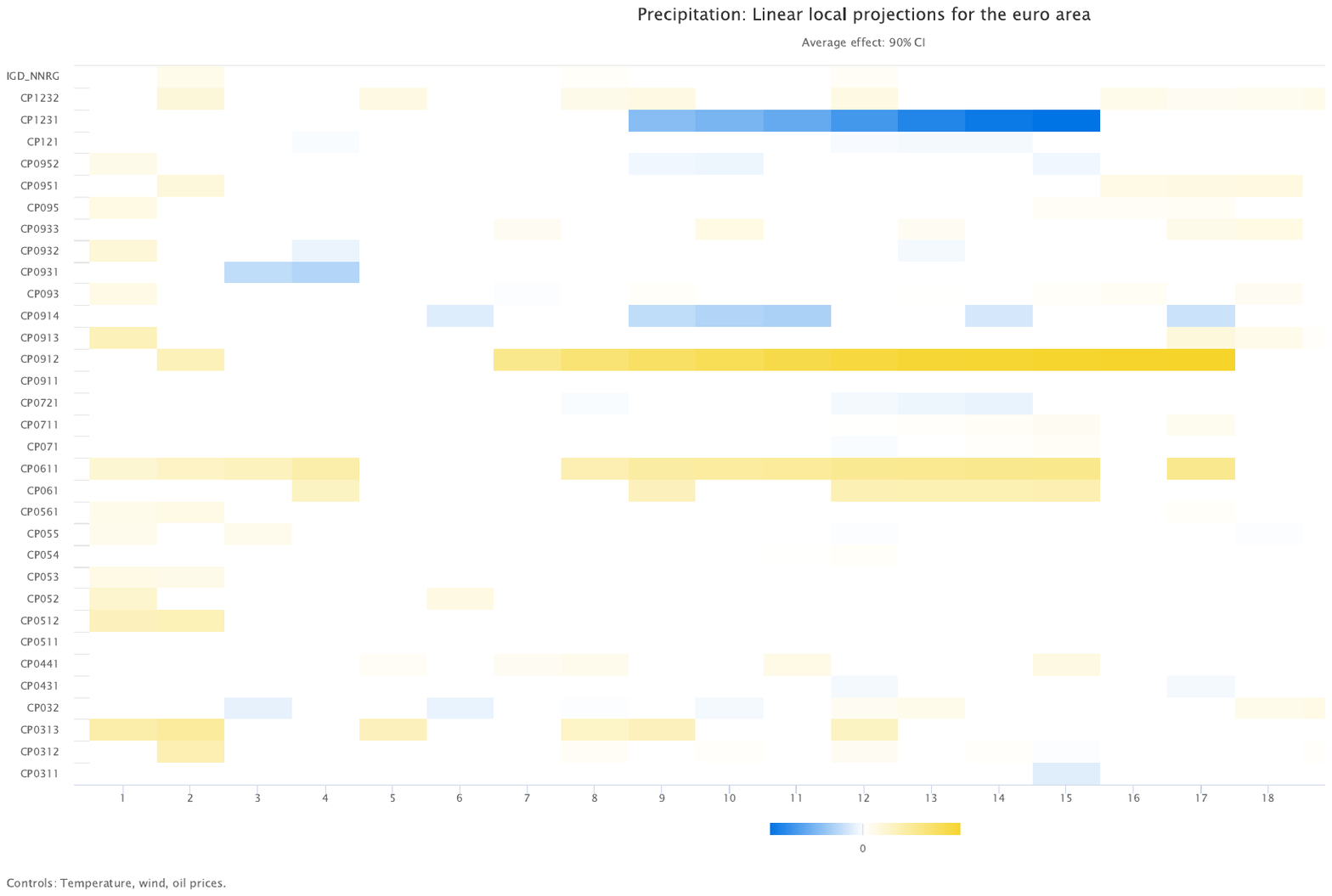}
	\end{subfigure}%
	~
	\begin{subfigure}[t]{0.8\textwidth}
		\centering
		\includegraphics[width=13cm, height=15cm]{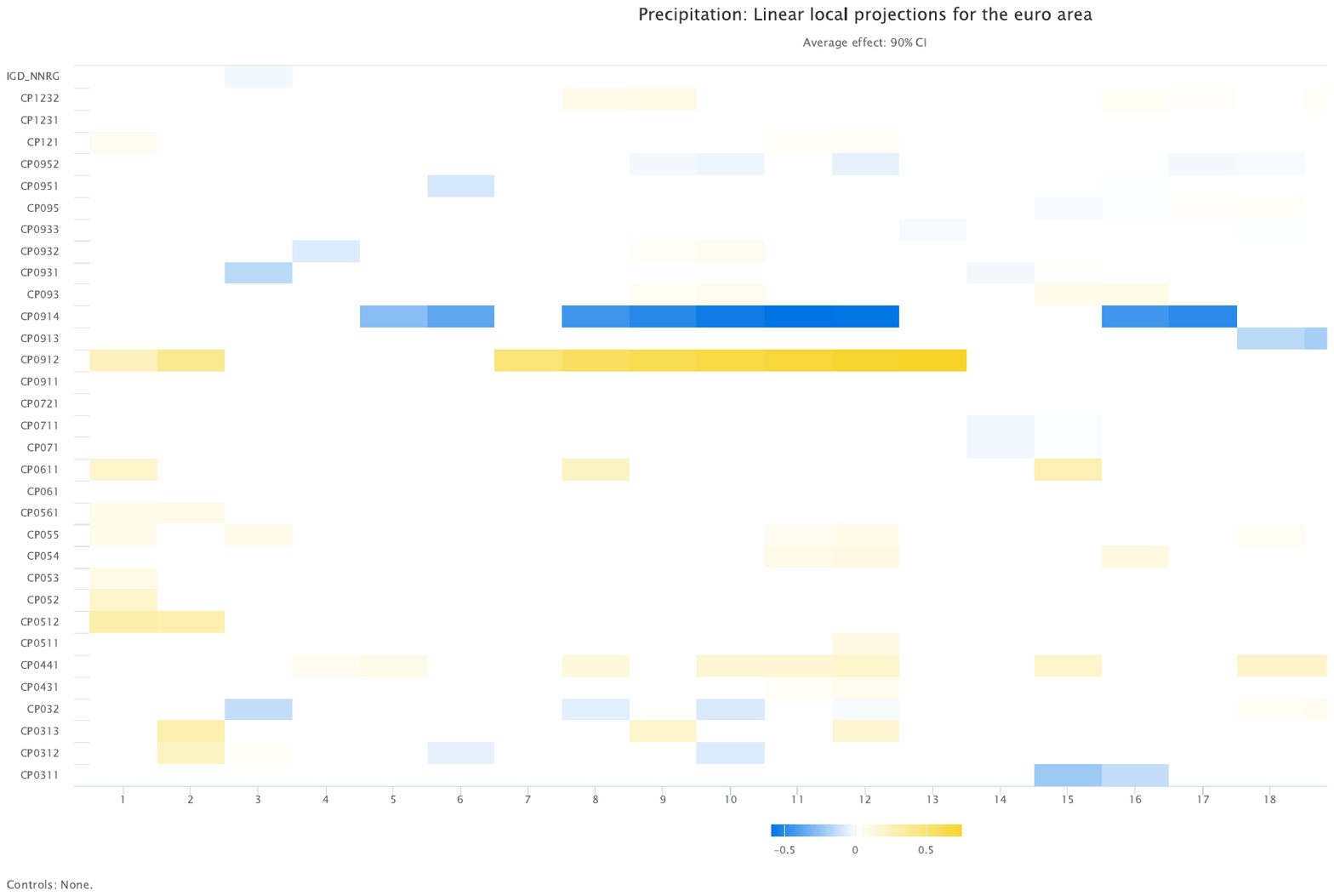}
	\end{subfigure}%
			\vspace{-0.1cm}
	\caption{Precipitation: Local projections for the Euro Area (Non-Energy industrial goods)}	
		 \label{fig:FigA2}
\end{figure*}	
\end{landscape}

\begin{landscape}
\captionsetup[subfigure]{aboveskip=1pt}
\begin{figure*}[th]
	\centering
	\begin{subfigure}[t]{0.8\textwidth}
		\centering
		\includegraphics[width=13cm, height=15cm]{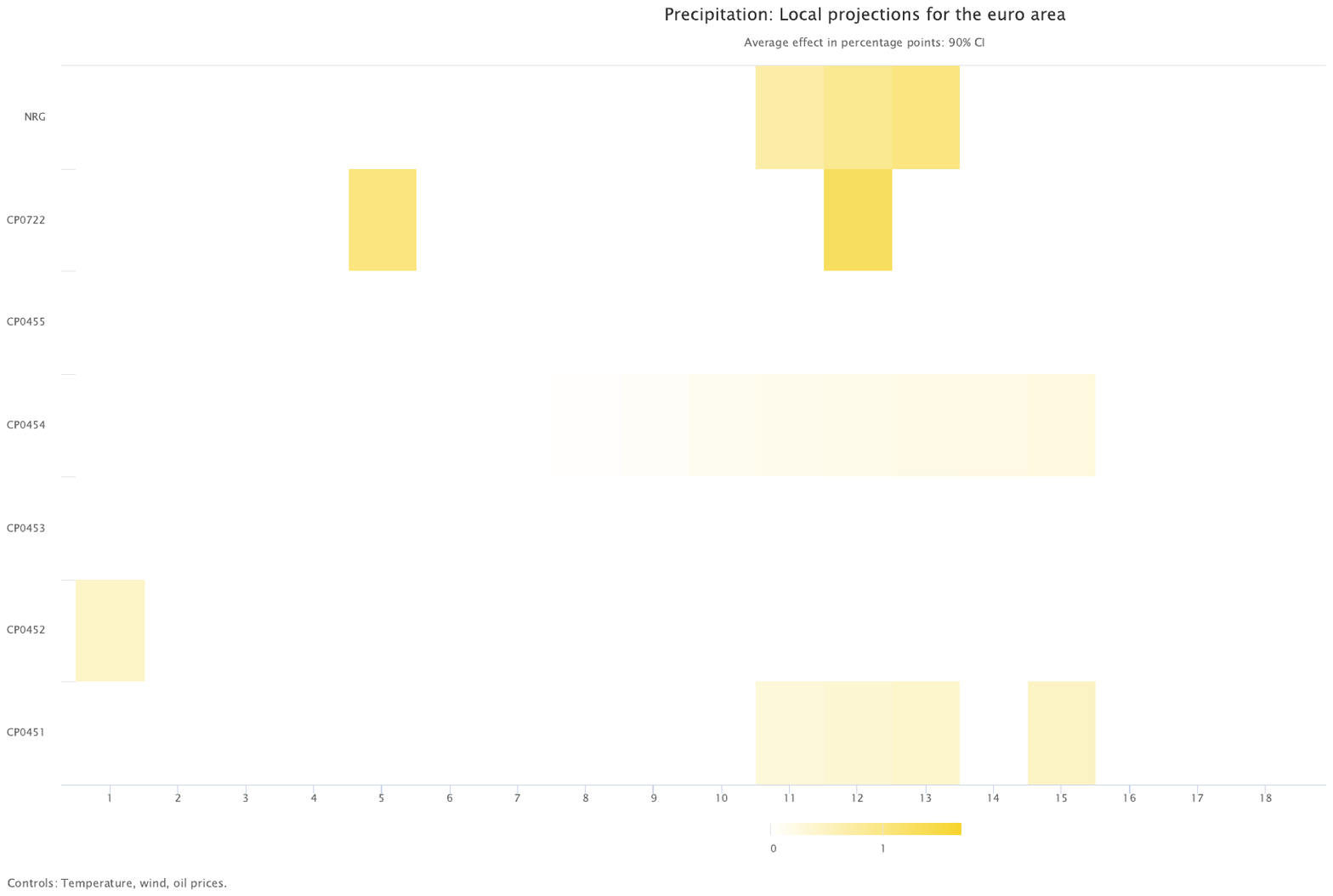}
	\end{subfigure}%
	~
	\begin{subfigure}[t]{0.8\textwidth}
		\centering
		\includegraphics[width=13cm, height=15cm]{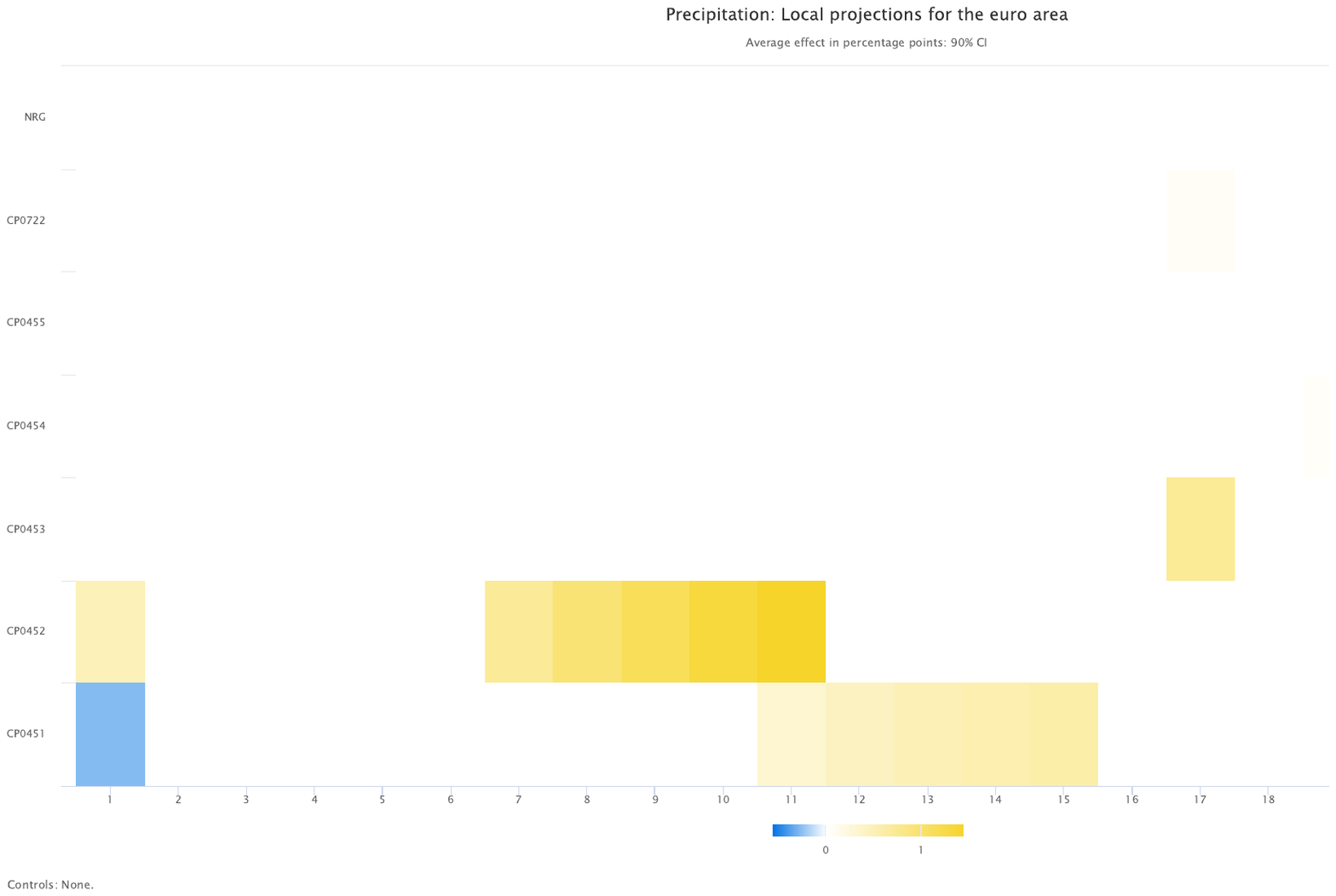}
	\end{subfigure}%
	\vspace{-0.1cm}

	\caption{Precipitation: Local projections for the Euro Area (Energy)}	
		 \label{fig:FigA3}
\end{figure*}	

\end{landscape}

\begin{landscape}
\captionsetup[subfigure]{aboveskip=1pt}
\begin{figure*}[th]
	\centering
	\begin{subfigure}[t]{0.8\textwidth}
		\centering
		\includegraphics[width=14cm, height=14cm]{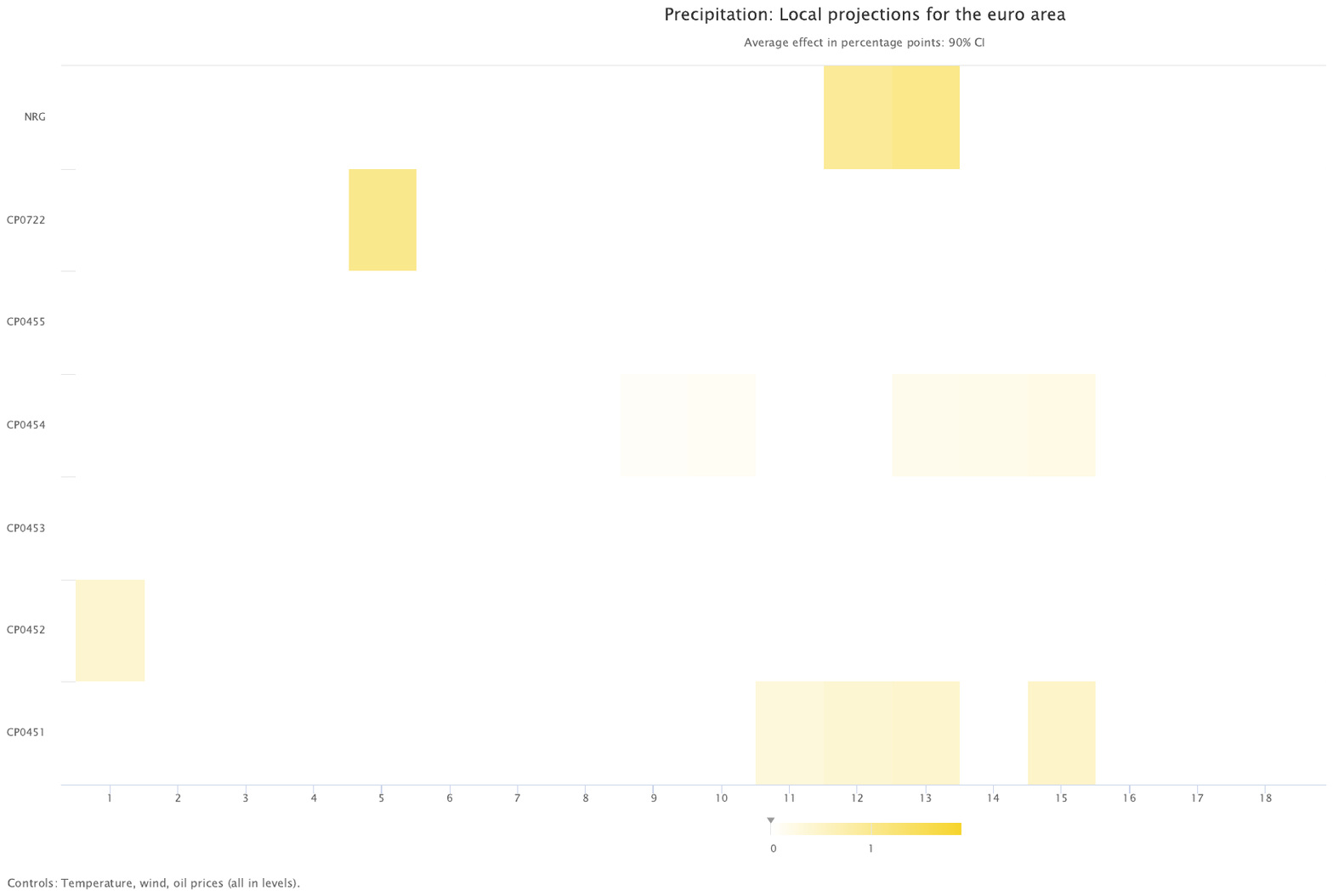}
	\end{subfigure}%
	\caption{Precipitation: Local projections for the Euro Area (Energy in levels)}	
		 \label{fig:FigA4}
\end{figure*}	
\end{landscape}

\begin{landscape}
\captionsetup[subfigure]{aboveskip=1pt}
\begin{figure*}[th]
	\centering
	\begin{subfigure}[t]{0.8\textwidth}
		\centering
		\includegraphics[width=13cm, height=15cm]{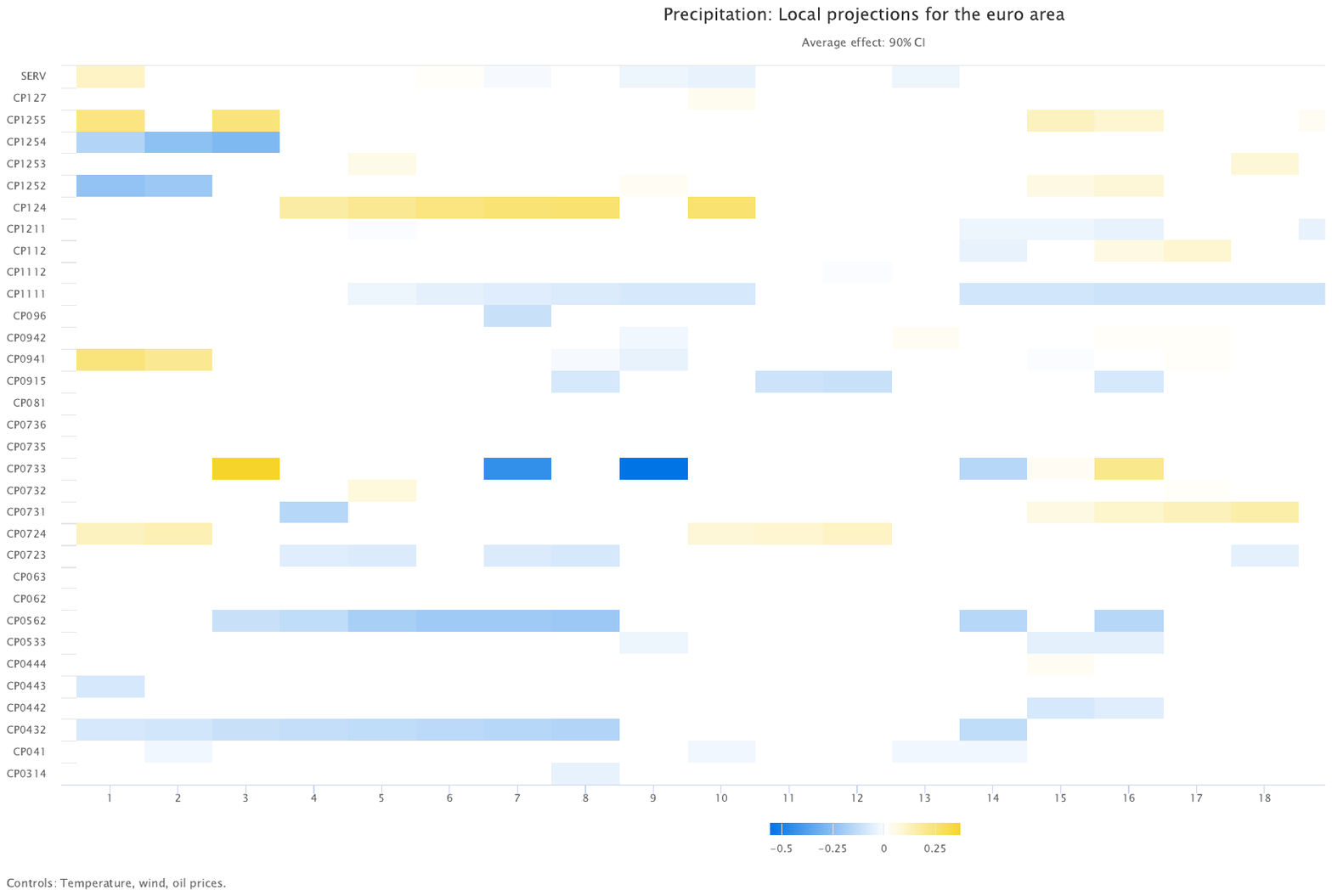}
	\end{subfigure}%
	~
	\begin{subfigure}[t]{0.8\textwidth}
		\centering
		\includegraphics[width=13cm, height=15cm]{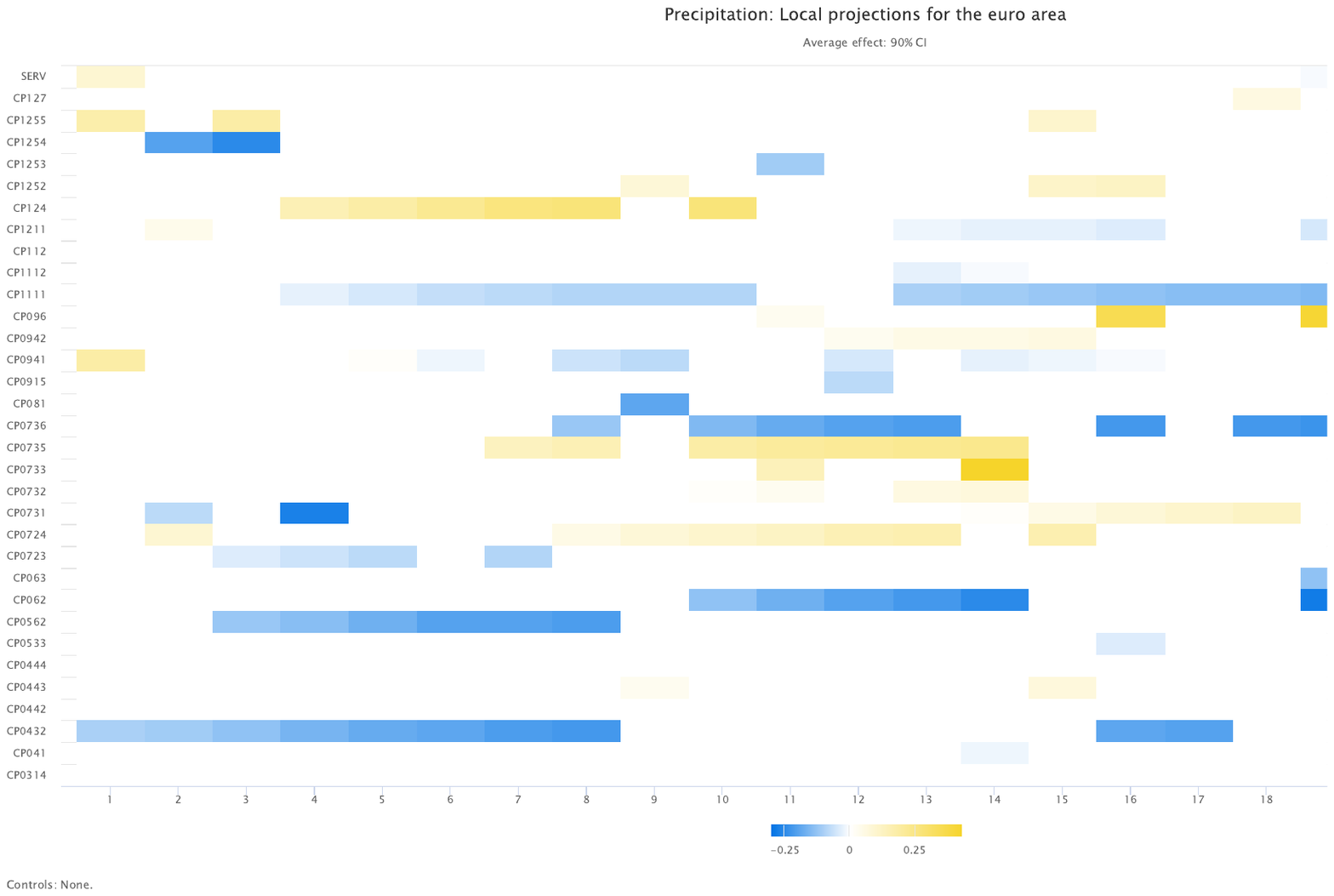}
	\end{subfigure}%
			\vspace{-0.1cm}
	\caption{Precipitation: Local projections for the Euro Area (Services)}	
		 \label{fig:FigA5}
\end{figure*}	
\end{landscape}

\begin{landscape}
\captionsetup[subfigure]{aboveskip=1pt}
\begin{figure*}[th]
	\centering
	\begin{subfigure}[t]{0.8\textwidth}
		\centering
		\includegraphics[width=13cm, height=15cm]{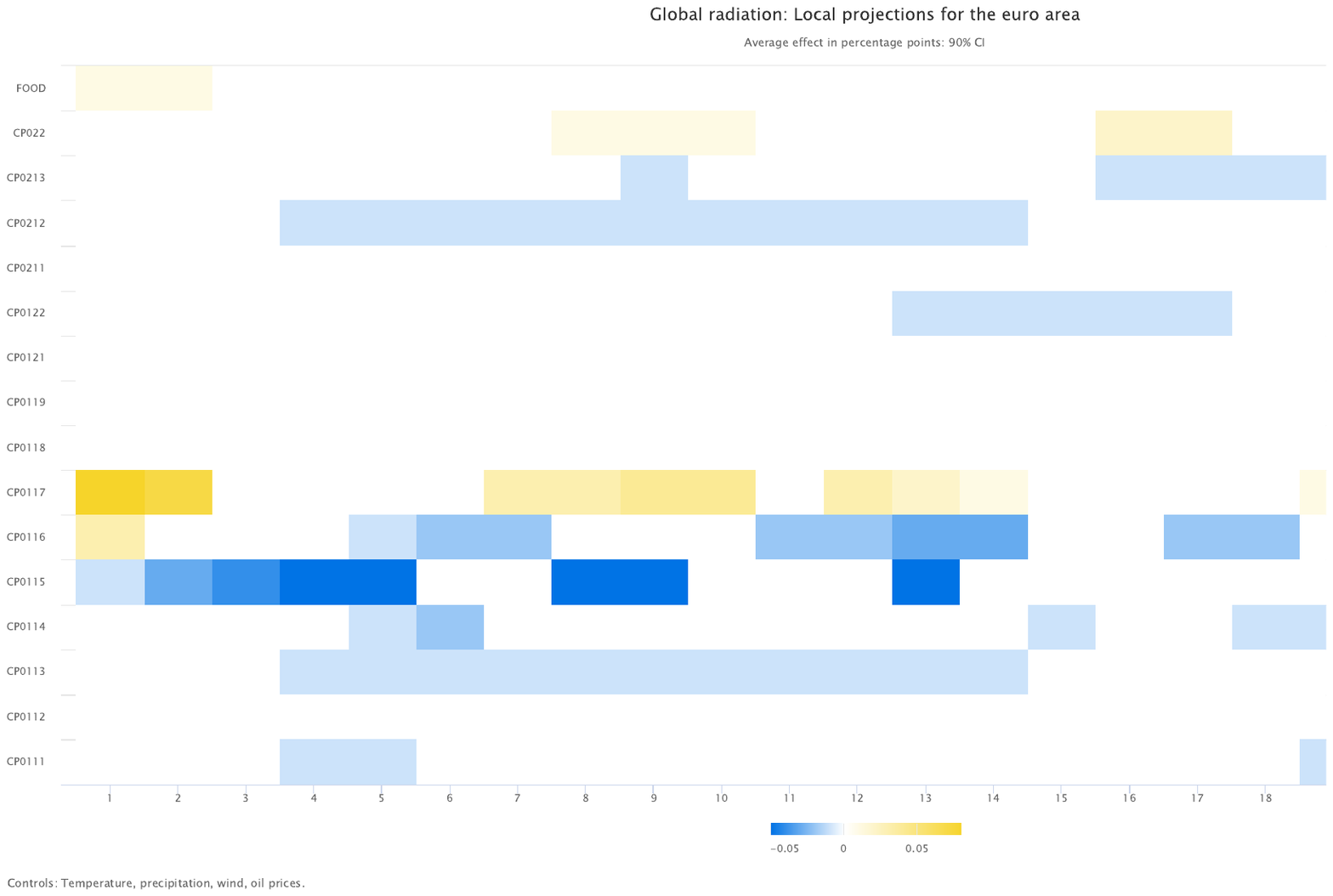}
	\end{subfigure}%
	~
	\begin{subfigure}[t]{0.8\textwidth}
		\centering
		\includegraphics[width=13cm, height=15cm]{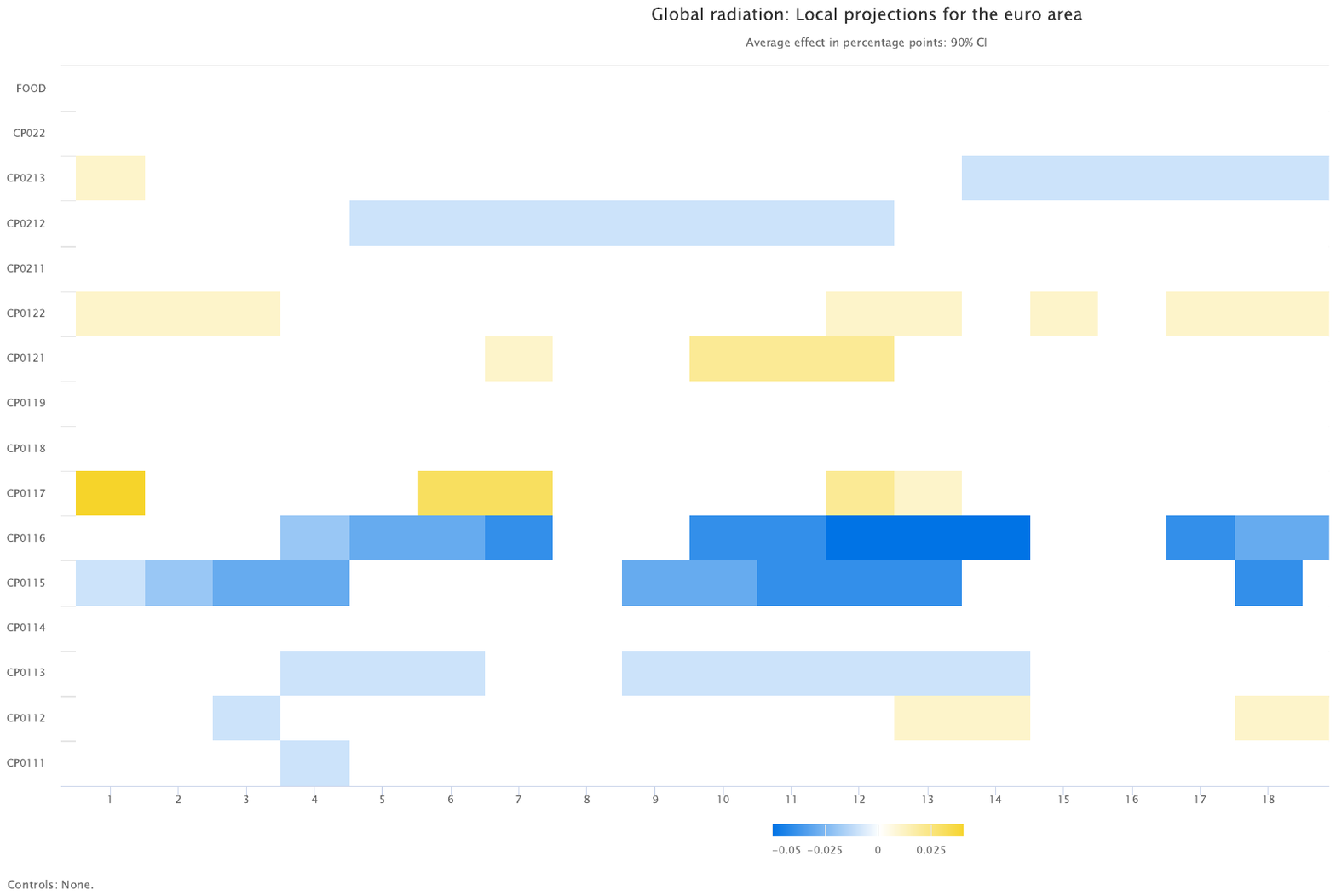}
	\end{subfigure}%
			\vspace{-0.1cm}
   	\caption{Global Radiation: Local projections for the Euro Area (Food)}	
		 \label{fig:FigA6}
\end{figure*}	

\end{landscape}

\begin{landscape}
\captionsetup[subfigure]{aboveskip=1pt}
\begin{figure*}[th]
	\centering
	\begin{subfigure}[t]{0.8\textwidth}
		\centering
	\includegraphics[width=13cm, height=15cm]{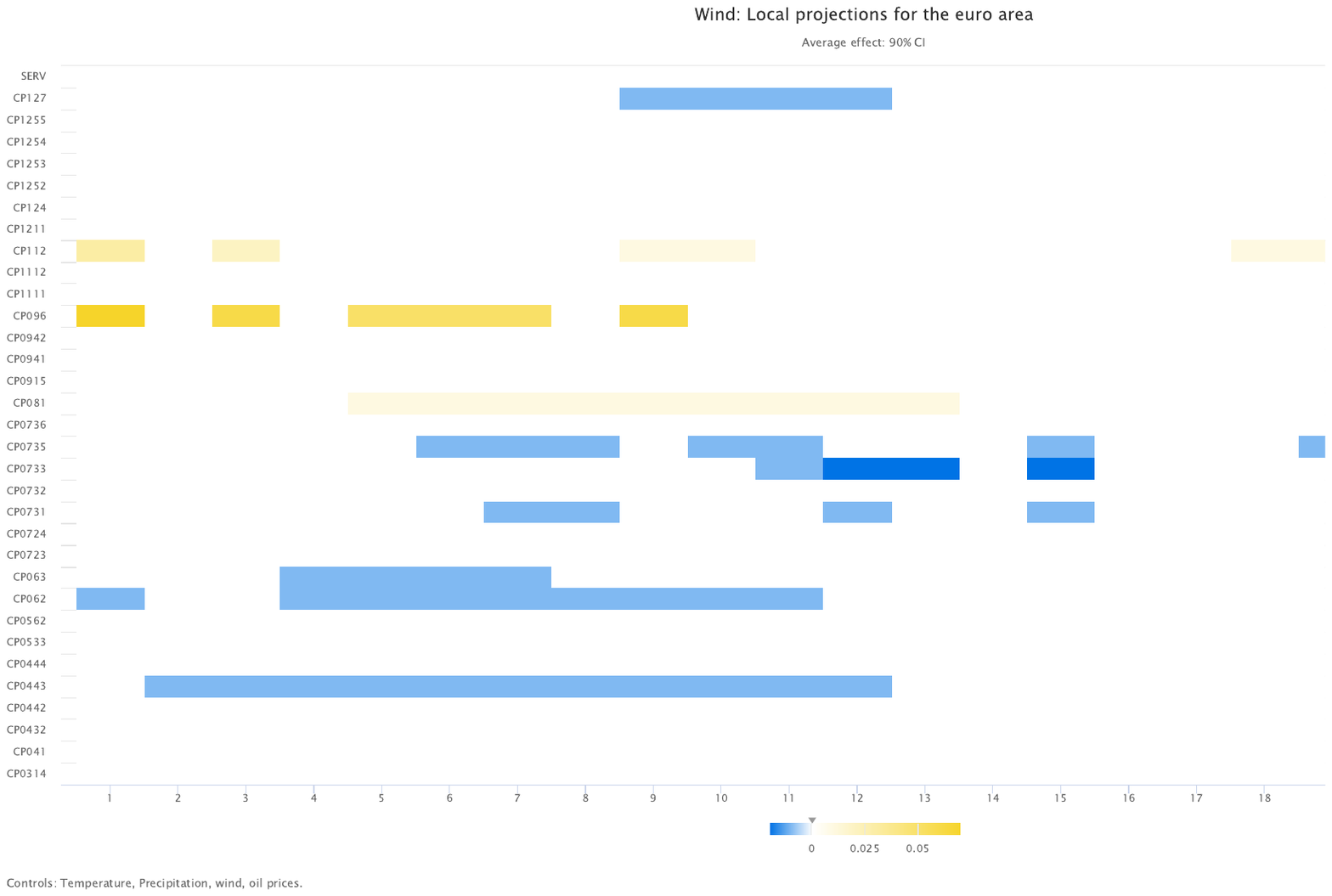}
	\end{subfigure}%
	~
	\begin{subfigure}[t]{0.8\textwidth}
		\centering
		\includegraphics[width=13cm, height=15cm]{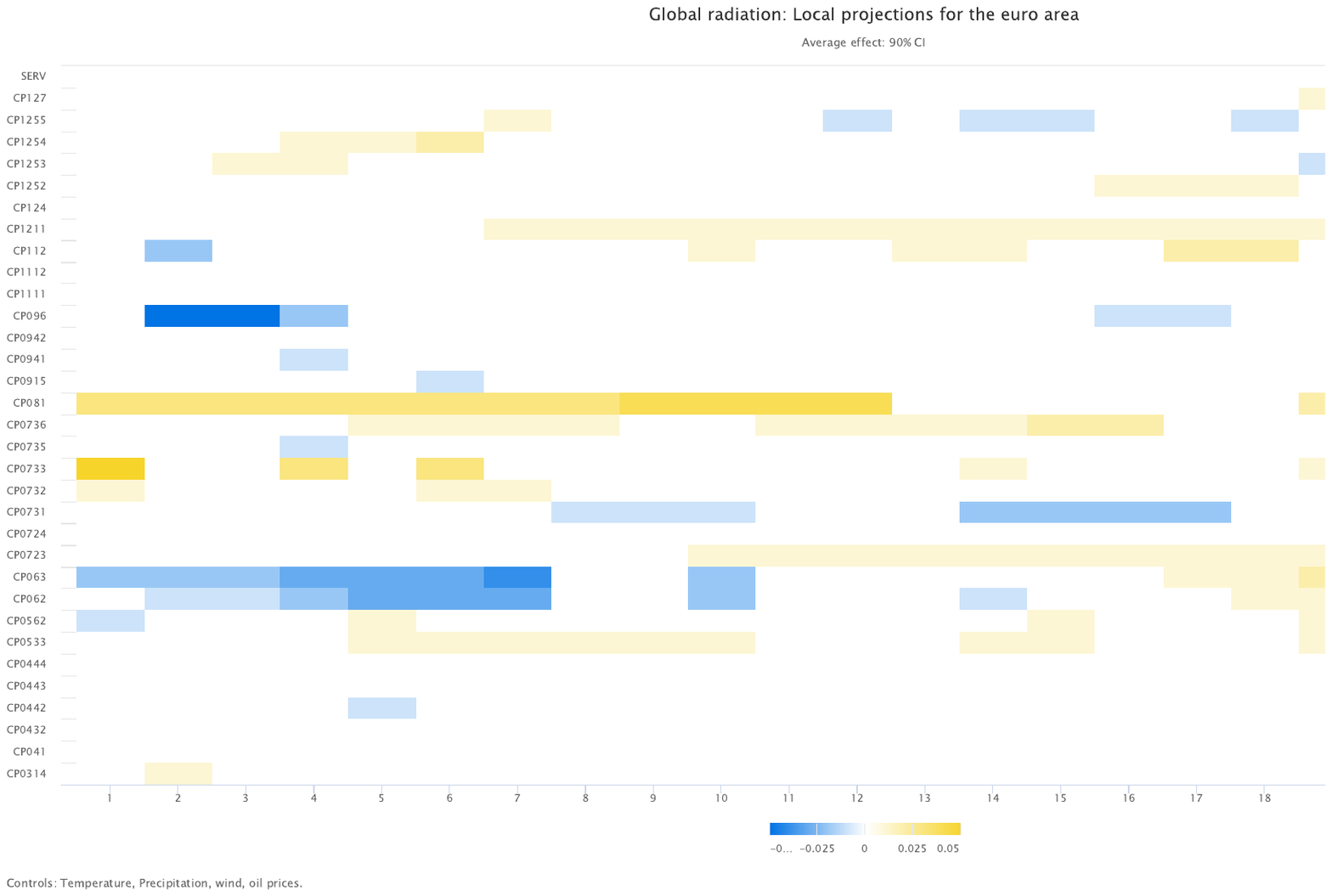}
	\end{subfigure}%
			\vspace{-0.1cm}
   	\caption{Solar Radiation: Local projections for the Euro Area (Services - max and min)}	
		 \label{fig:FigA7}
\end{figure*}	

\end{landscape}

\begin{landscape}
\captionsetup[subfigure]{aboveskip=1pt}
\begin{figure*}[th]
	\centering
	\begin{subfigure}[t]{0.8\textwidth}
		\centering
		\includegraphics[width=13cm, height=15cm]{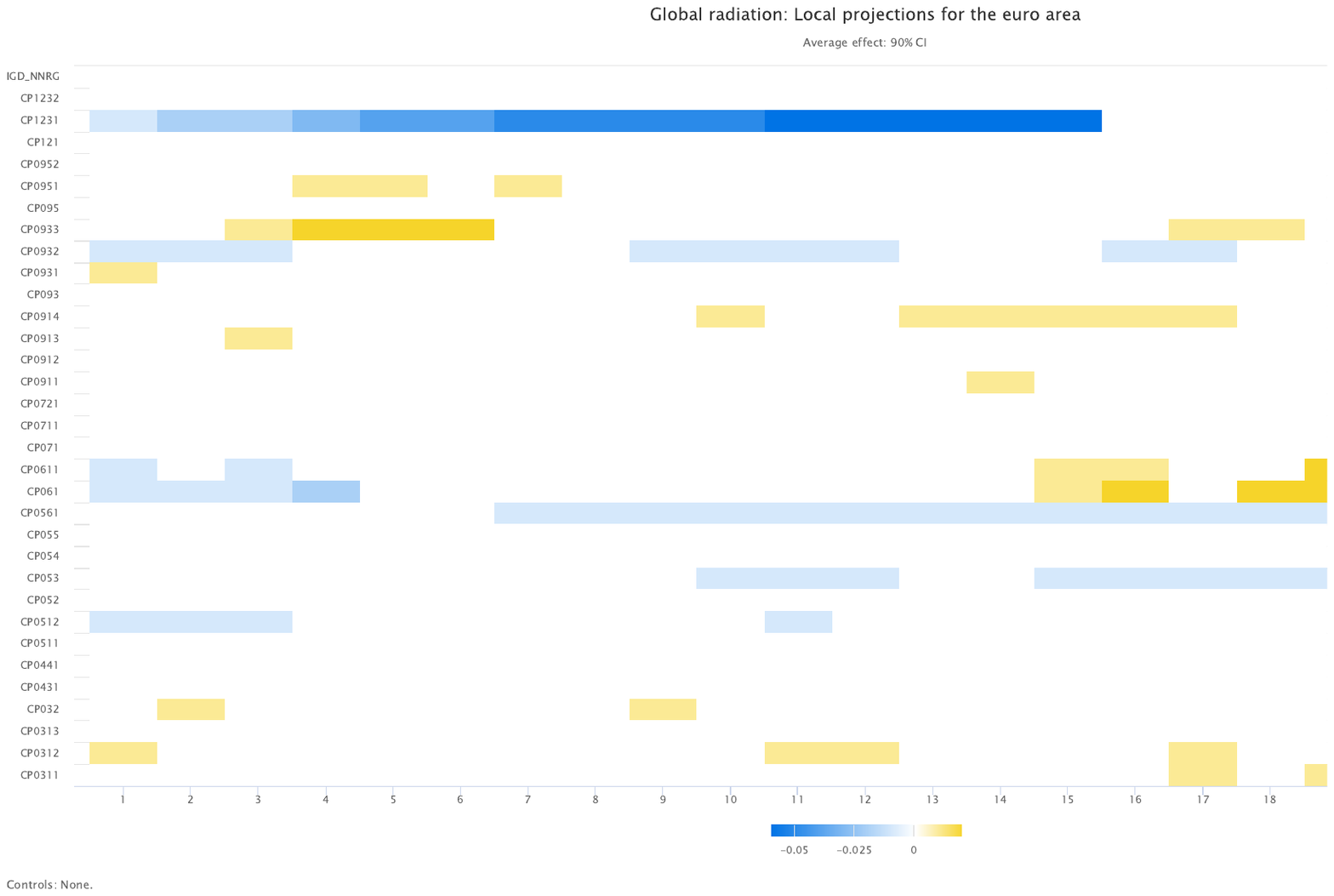}
	\end{subfigure}%
  	\caption{Global Radiation: Local projections for the Euro Area (Non-Energy industrial goods)}	
		 \label{fig:FigA8}
\end{figure*}	

\end{landscape}

\begin{landscape}
\captionsetup[subfigure]{aboveskip=1pt}
\begin{figure*}[th]
	\centering
	\begin{subfigure}[t]{0.8\textwidth}
		\centering
		\includegraphics[width=13cm, height=15cm]{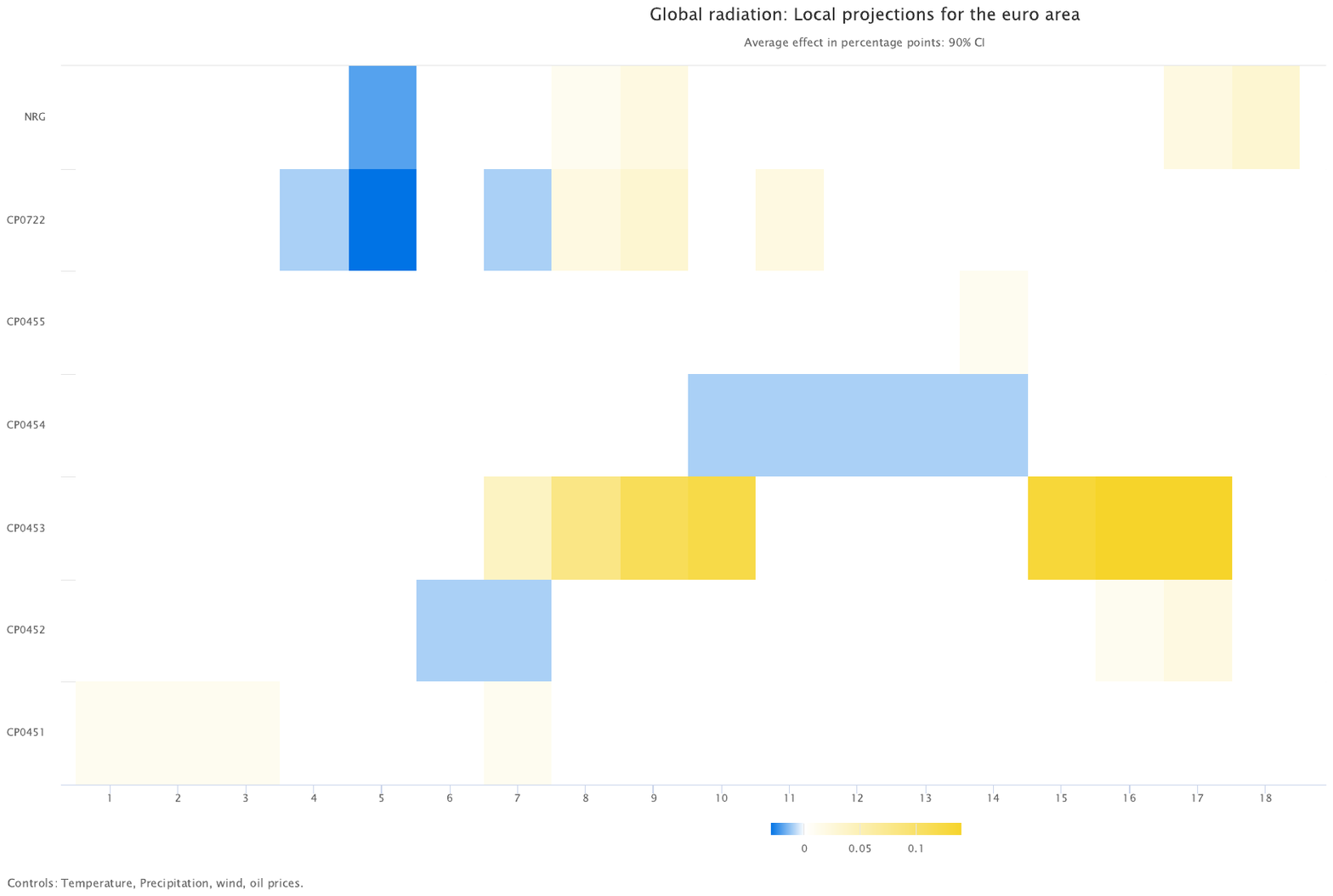}
	\end{subfigure}%
	~
	\begin{subfigure}[t]{0.8\textwidth}
		\centering
		\includegraphics[width=13cm, height=15cm]{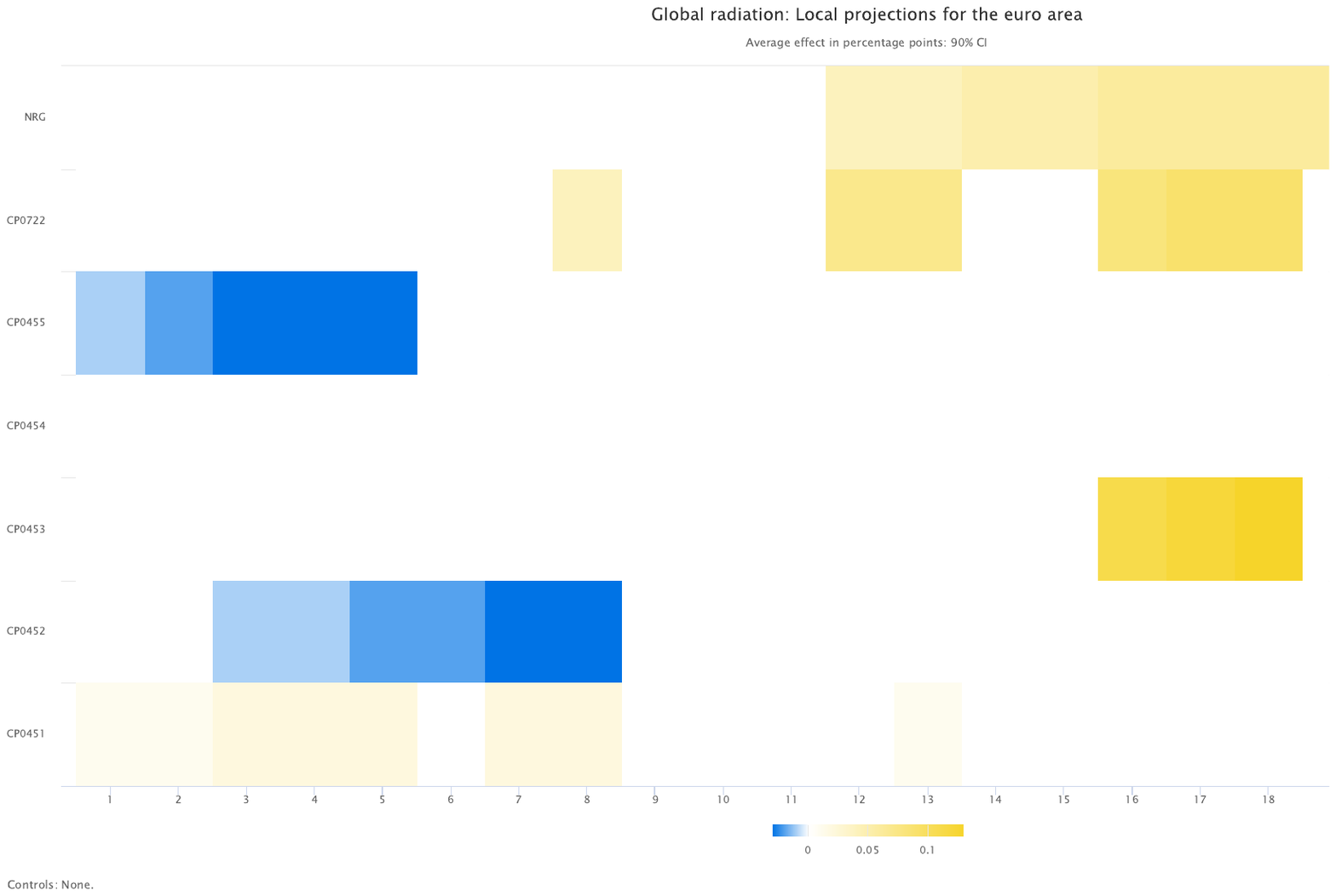}
	\end{subfigure}%
     \vspace{-0.1cm}
  	\caption{Global Radiation: Local projections for the Euro Area (Energy)}			
		 \label{fig:FigA9}
\end{figure*}	
\end{landscape}

\begin{landscape}
\captionsetup[subfigure]{aboveskip=1pt}
\begin{figure*}[th]
	\centering
	\begin{subfigure}[t]{0.8\textwidth}
		\centering
		\includegraphics[width=13cm, height=15cm]{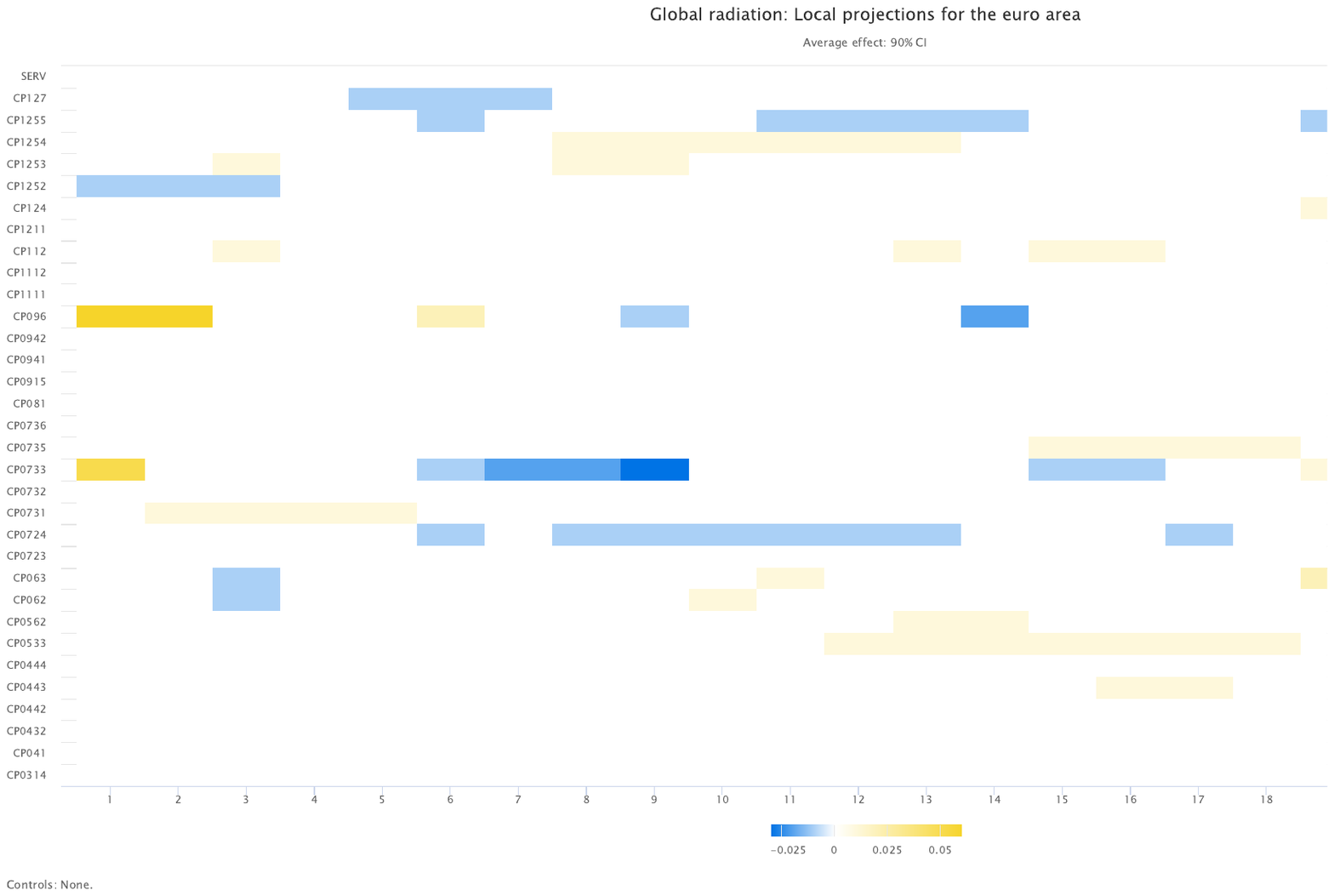}
	\end{subfigure}%
\caption{Global Radiation: Local projections for the Euro Area (Services)}		
		 \label{fig:FigA10}
\end{figure*}	
\end{landscape}

\begin{landscape}
\captionsetup[subfigure]{aboveskip=1pt}
\begin{figure*}[th]
	\centering
	\begin{subfigure}[t]{0.8\textwidth}
		\centering
		\includegraphics[width=13cm, height=15cm]{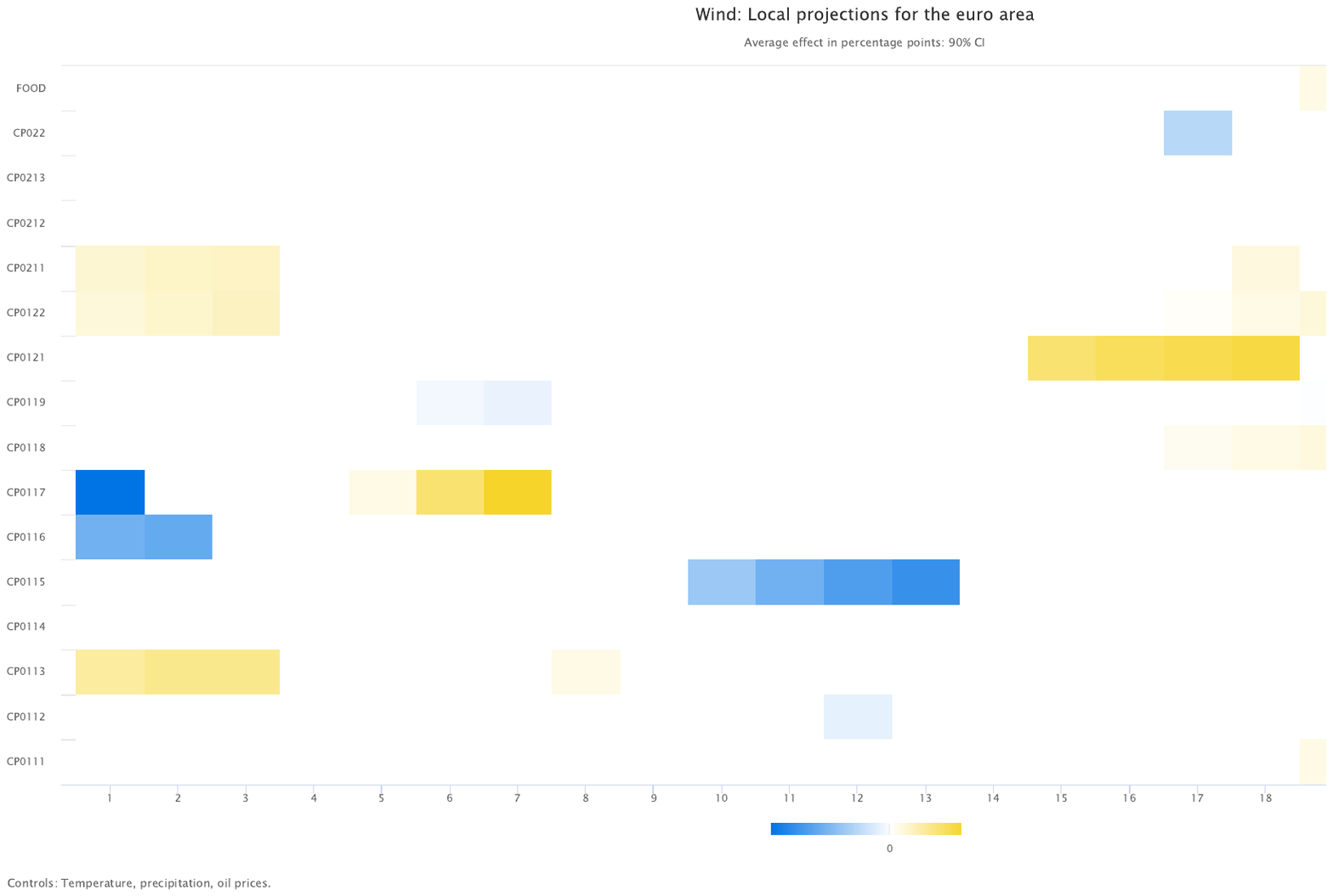}
	\end{subfigure}%
\caption{Wind: Local projections for the Euro Area (Food)}		
		 \label{fig:FigA11}
\end{figure*}	

\end{landscape}

\begin{landscape}
\captionsetup[subfigure]{aboveskip=1pt}
\begin{figure*}[th]
	\centering
	\begin{subfigure}[t]{0.8\textwidth}
		\centering
		\includegraphics[width=13cm, height=15cm]{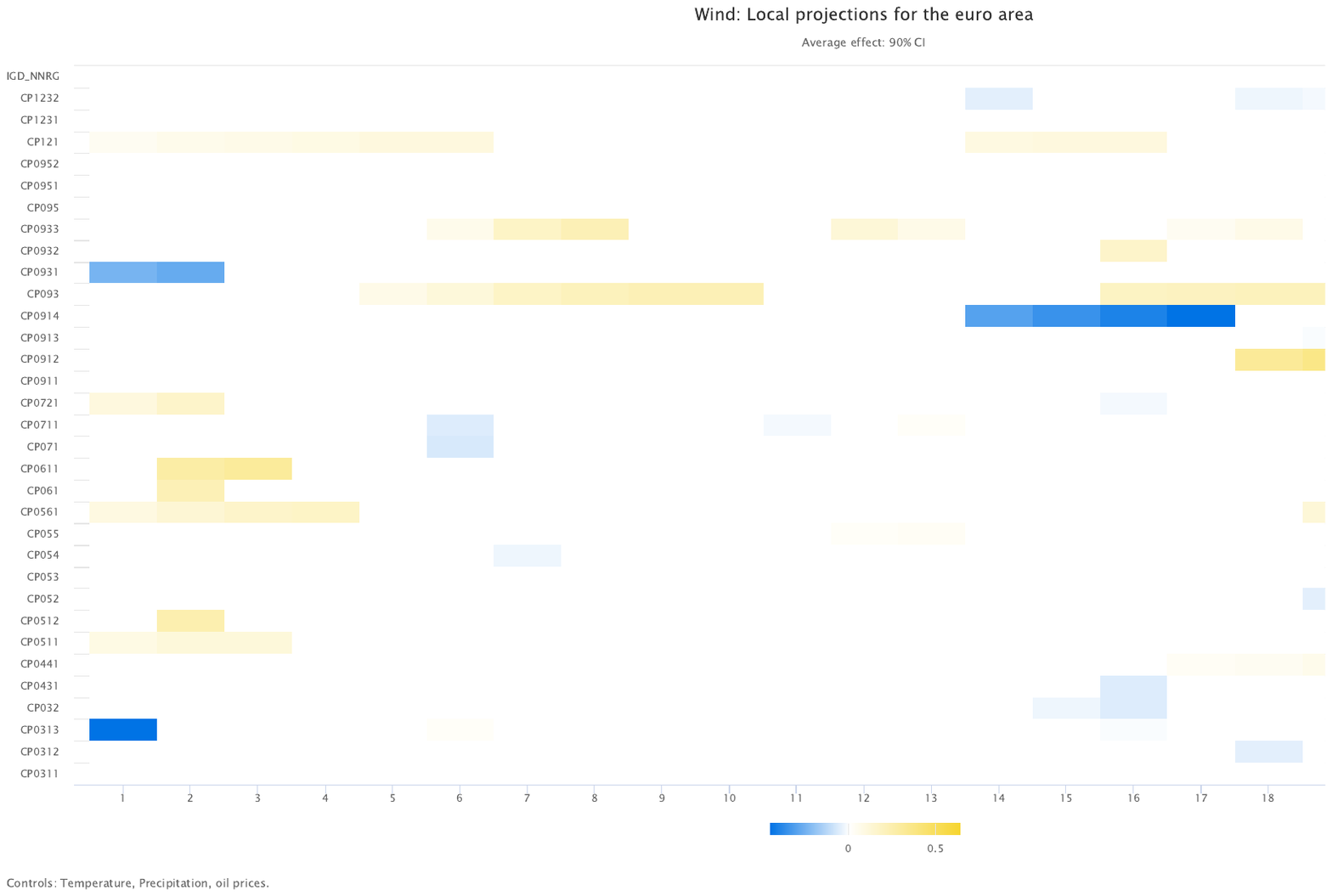}
	\end{subfigure}%
\caption{Wind: Local projections for the Euro Area (Non-Energy industrial goods)}		
		 \label{fig:FigA12}
\end{figure*}	

\end{landscape}

\begin{landscape}
\captionsetup[subfigure]{aboveskip=1pt}
\begin{figure*}[th]
	\centering
	\begin{subfigure}[t]{0.8\textwidth}
		\centering
		\includegraphics[width=13cm, height=15cm]{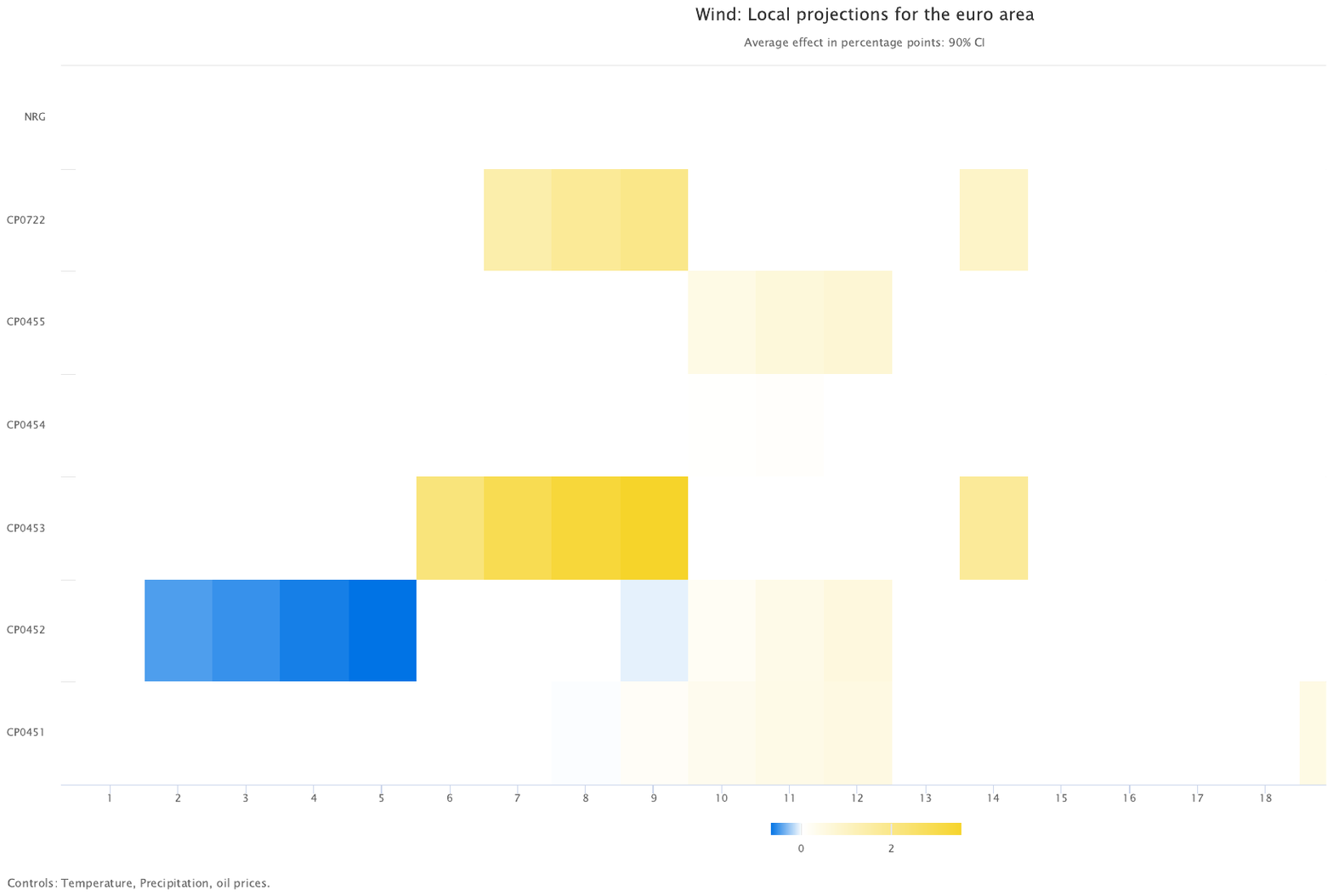}
	\end{subfigure}%
\caption{Wind: Local projections for the Euro Area (Energy)}		
		 \label{fig:FigA13}
\end{figure*}	

\end{landscape}

\begin{landscape}
\captionsetup[subfigure]{aboveskip=1pt}
\begin{figure*}[th]
	\centering
	\begin{subfigure}[t]{0.8\textwidth}
		\centering
		\includegraphics[width=13cm, height=15cm]{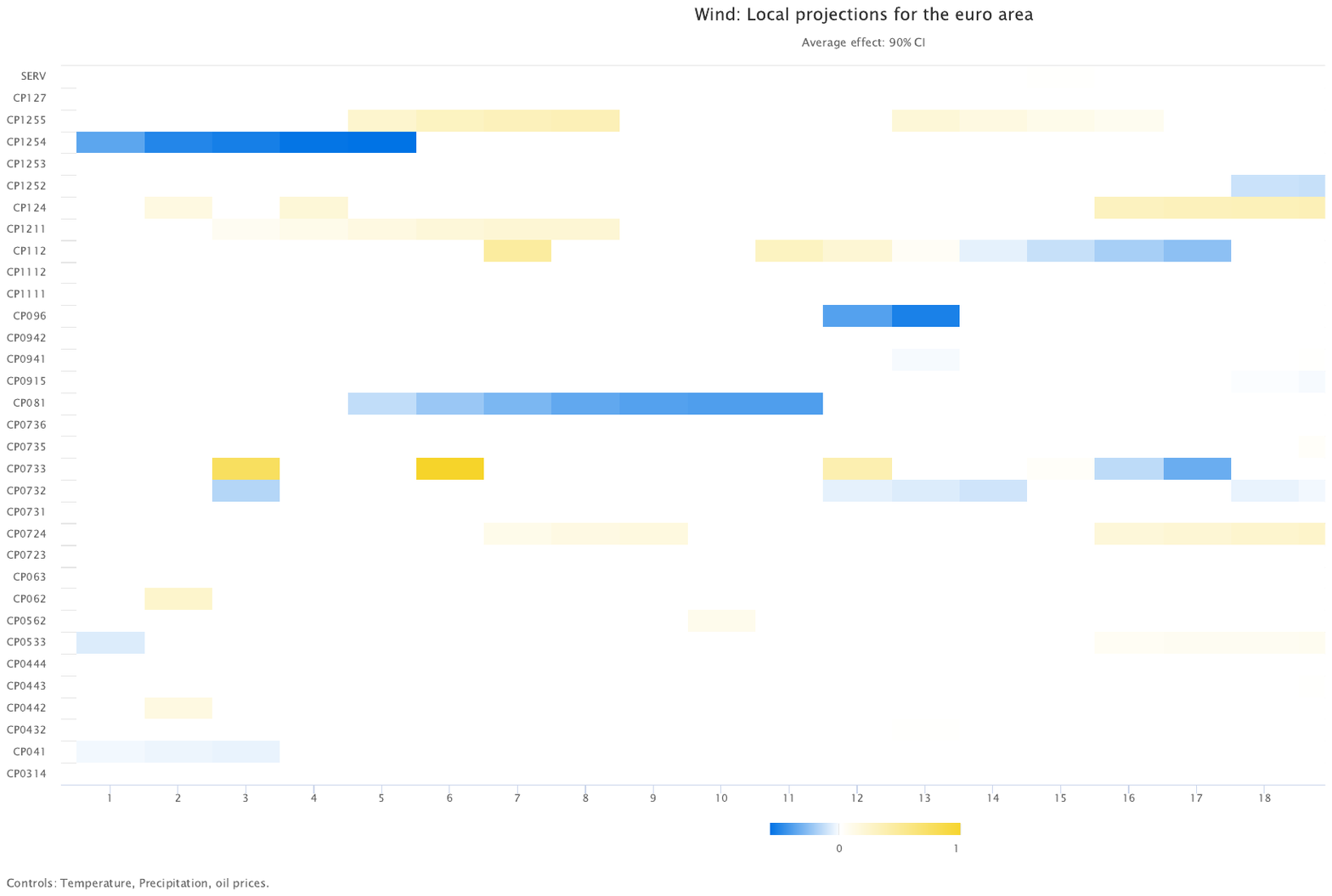}
	\end{subfigure}%
	~
	\begin{subfigure}[t]{0.8\textwidth}
		\centering
		\includegraphics[width=13cm, height=15cm]{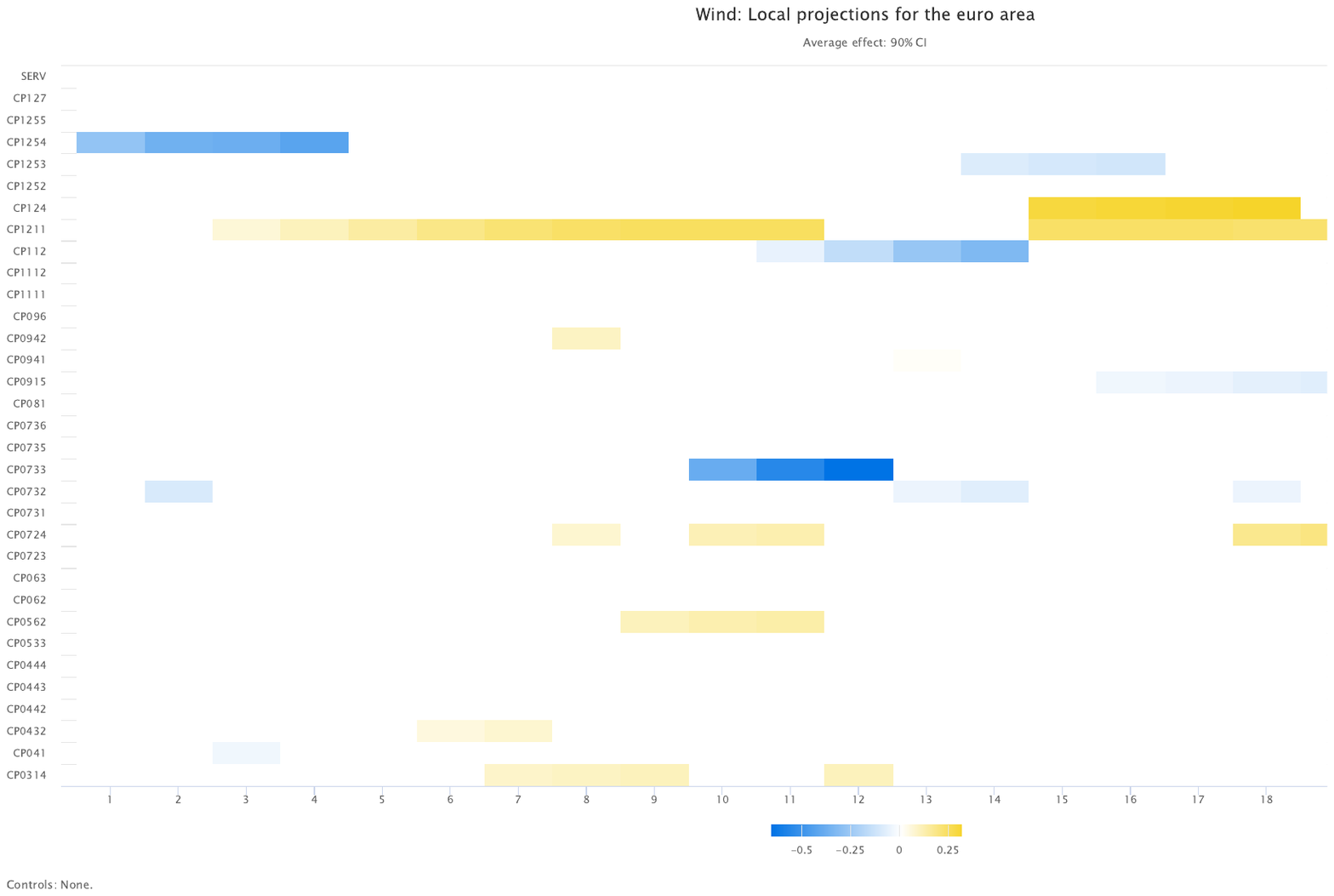}
	\end{subfigure}%
\caption{Wind: Local projections for the Euro Area (Services)}		
		 \label{fig:FigA14}
\end{figure*}	

\end{landscape}

\begin{landscape}
\captionsetup[subfigure]{aboveskip=1pt}
\begin{figure*}[th]
	\centering
	\begin{subfigure}[t]{0.8\textwidth}
		\centering
		\includegraphics[width=13cm, height=15cm]{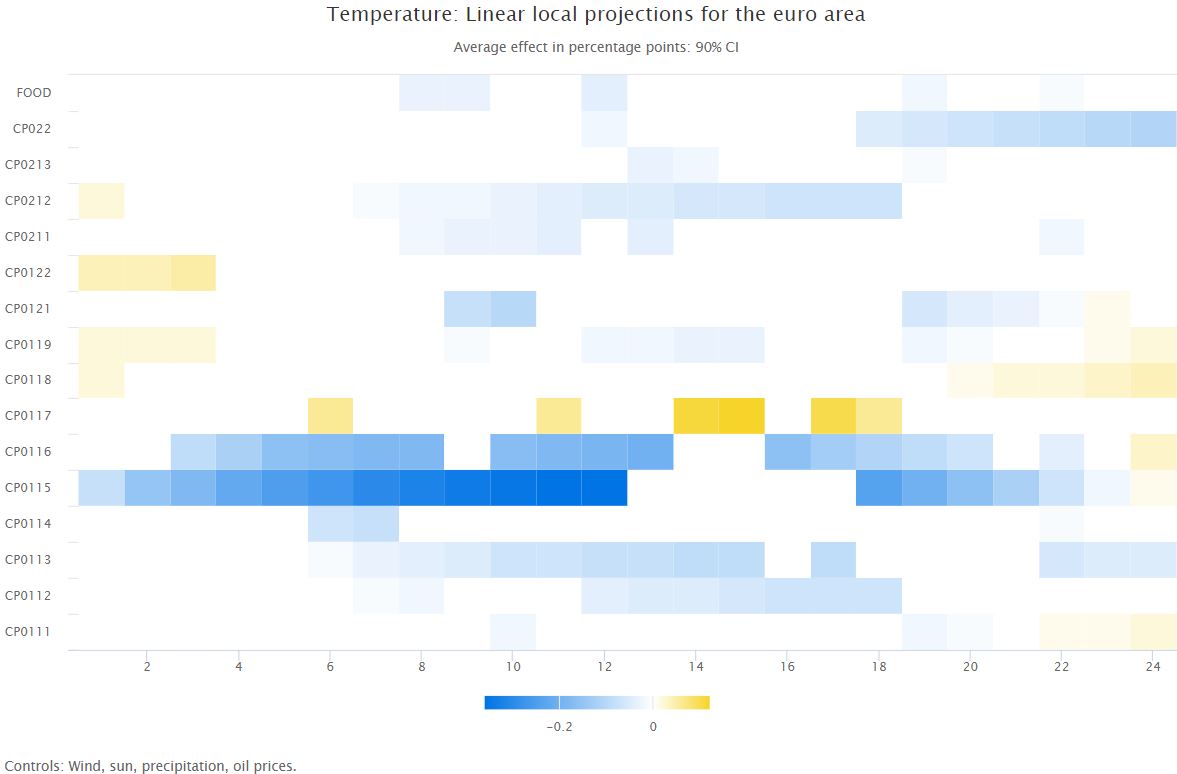}
	\end{subfigure}%
	~
	\begin{subfigure}[t]{0.8\textwidth}
		\centering
		\includegraphics[width=13cm, height=15cm]{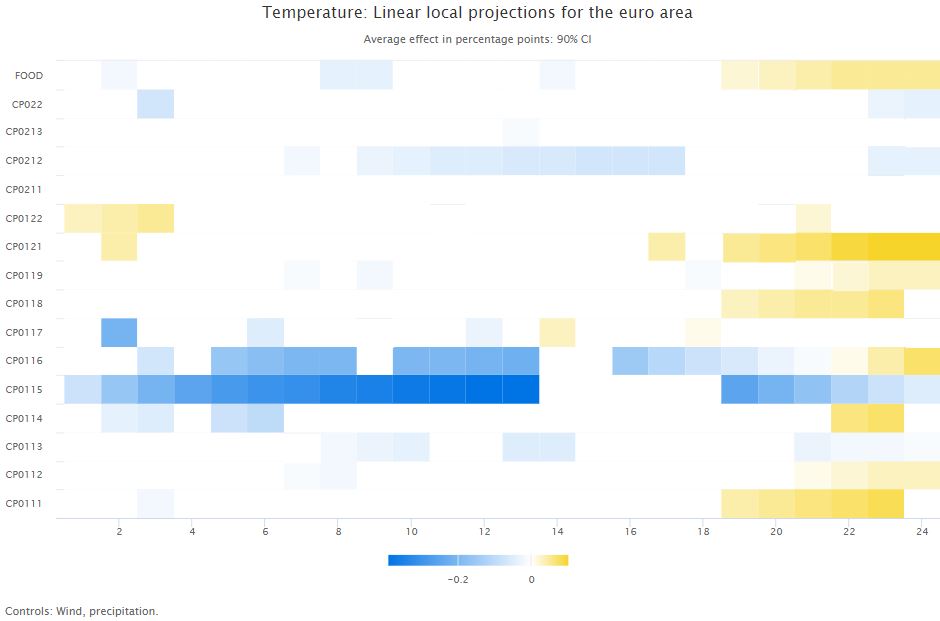}
	\end{subfigure}%
	
\caption{Temperature: Local projections for the Euro Area (Food)}		
		 \label{fig:FigA15}
\end{figure*}	

\end{landscape}

\begin{landscape}
\captionsetup[subfigure]{aboveskip=1pt}
\begin{figure*}[th]
	\centering
	\begin{subfigure}[t]{0.8\textwidth}
		\centering
		\includegraphics[width=13cm, height=15cm]{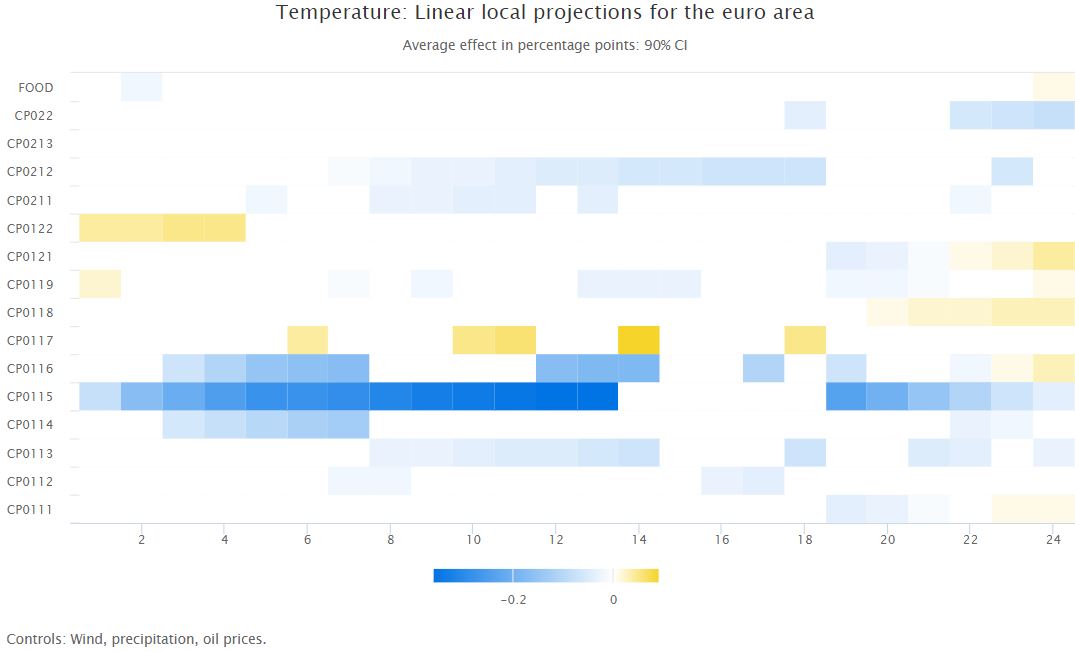}
	\end{subfigure}%
	~
	\begin{subfigure}[t]{0.8\textwidth}
		\centering
		\includegraphics[width=13cm, height=15cm]{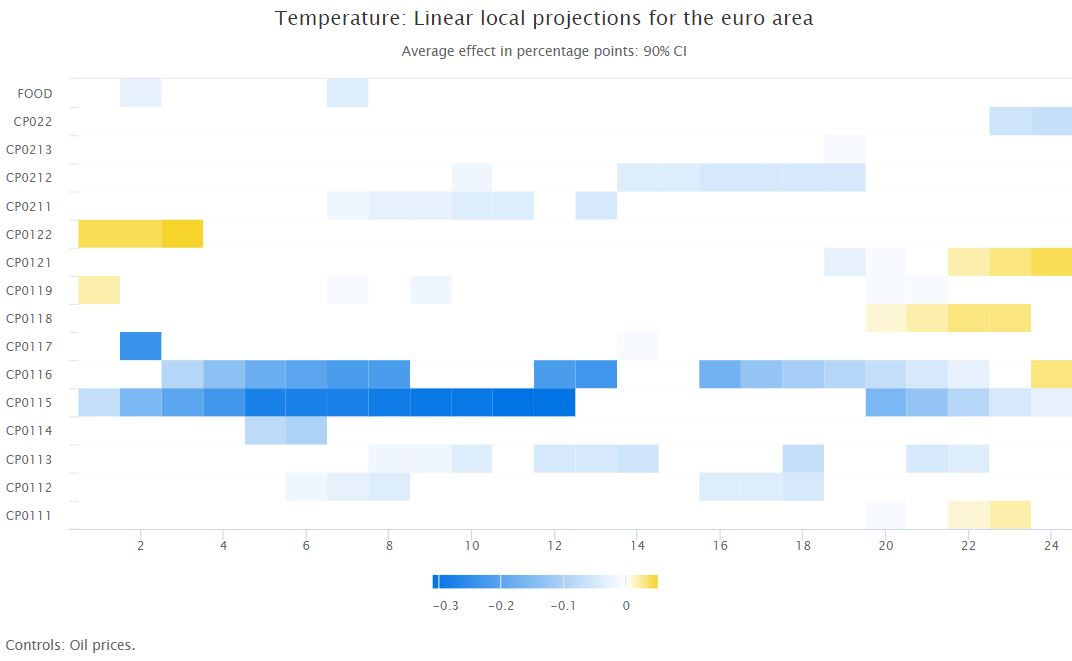}
	\end{subfigure}%
	\caption{Temperature: Local projections for the Euro Area (Food)}		
		 \label{fig:FigA16}
\end{figure*}	

\end{landscape}

\begin{landscape}
\captionsetup[subfigure]{aboveskip=1pt}
\begin{figure*}[th]
	\centering
	\begin{subfigure}[t]{0.8\textwidth}
		\centering
		\includegraphics[width=13cm, height=15cm]{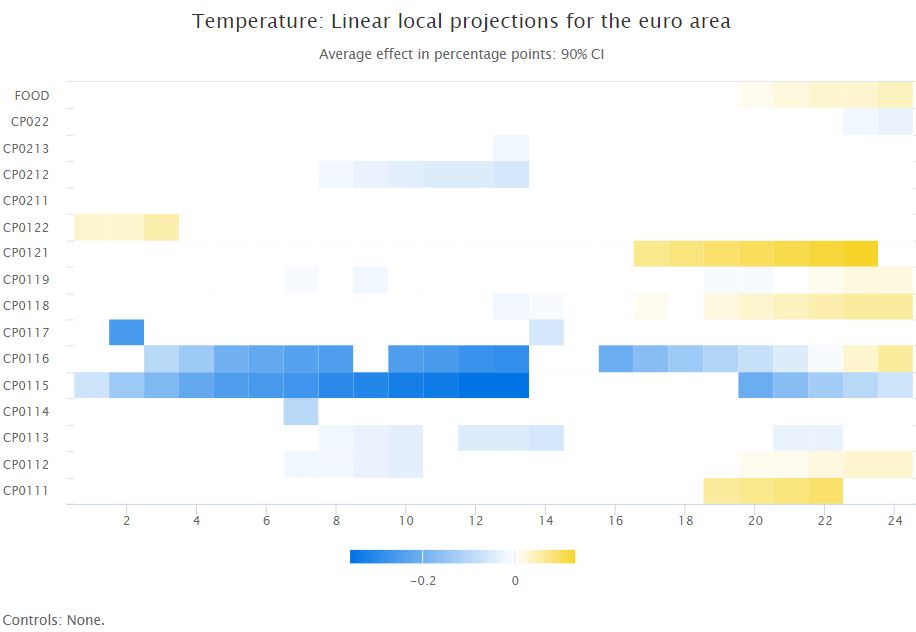}
	\end{subfigure}%
	~
	\begin{subfigure}[t]{0.8\textwidth}
		\centering
		\includegraphics[width=13cm, height=15cm]{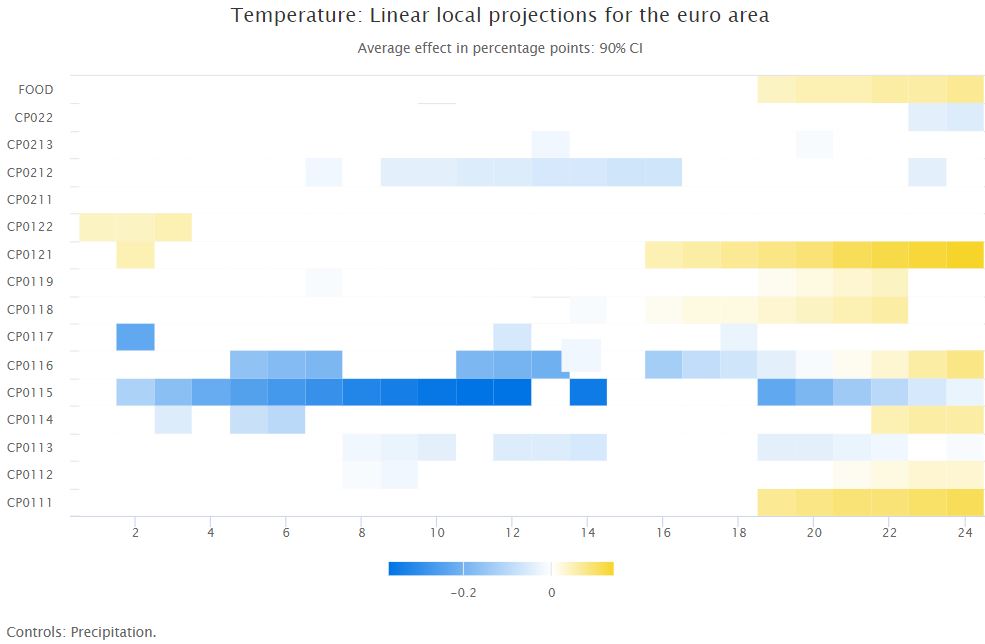}
	\end{subfigure}%
	\caption{Temperature: Local projections for the Euro Area (Food)}		
		 \label{fig:FigA17}
\end{figure*}	

\end{landscape}

\begin{landscape}
\captionsetup[subfigure]{aboveskip=1pt}
\begin{figure*}[th]
	\centering
	\begin{subfigure}[t]{0.8\textwidth}
		\centering
		\includegraphics[width=13cm, height=15cm]{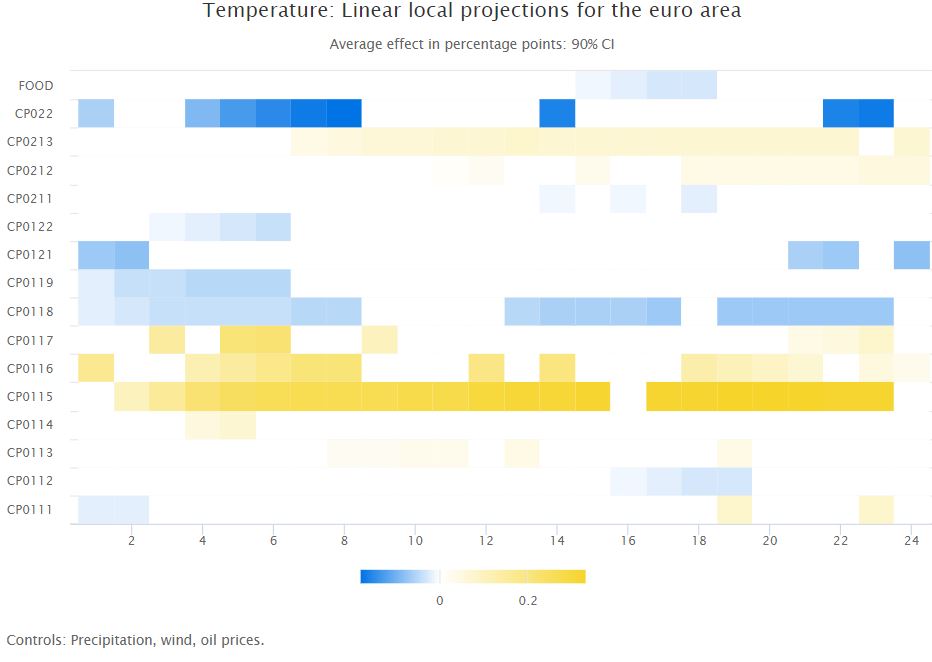}
	\end{subfigure}%
	~
	\begin{subfigure}[t]{0.8\textwidth}
		\centering
		\includegraphics[width=13cm, height=15cm]{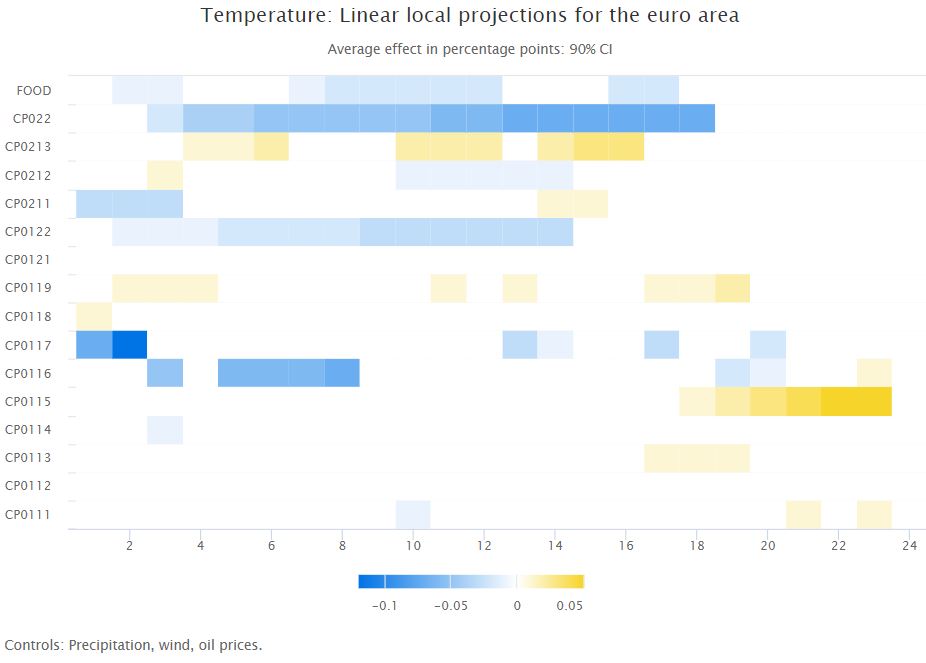}
	\end{subfigure}%
	\caption{Temperature: Local projections for the Euro Area (Food - Max and Min)}		
		 \label{fig:FigA18}	
\end{figure*}	

\end{landscape}

\begin{landscape}
\captionsetup[subfigure]{aboveskip=1pt}
\begin{figure*}[th]
	\centering
	\begin{subfigure}[t]{0.8\textwidth}
		\centering
		\includegraphics[width=13cm, height=15cm]{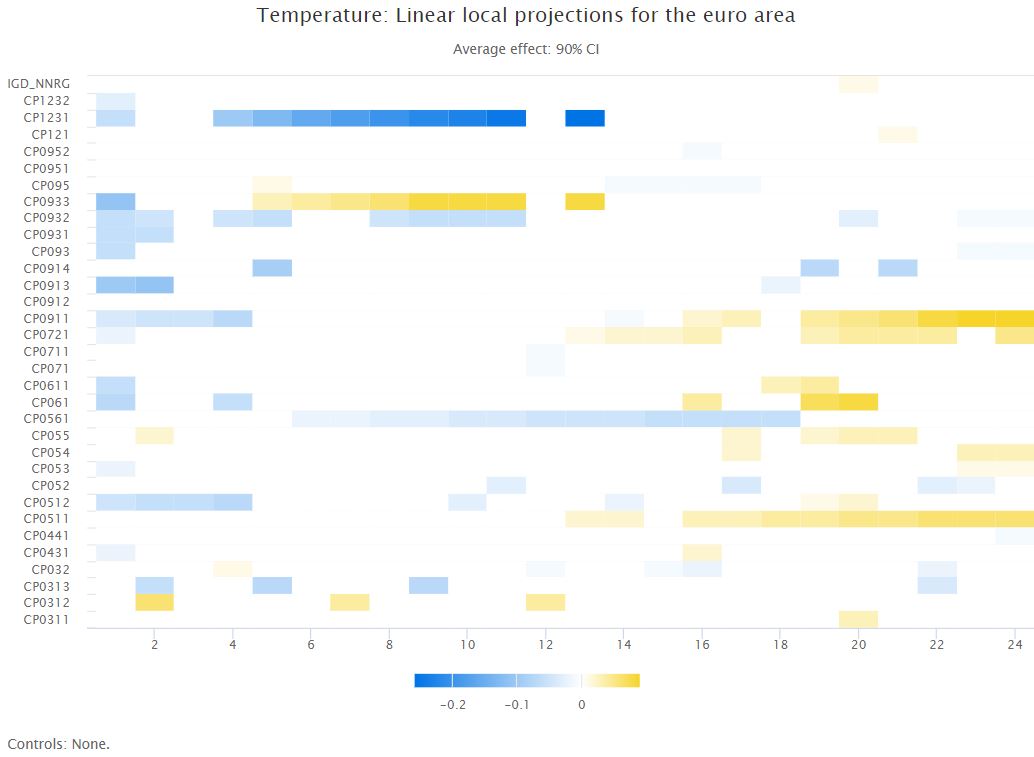}
	\end{subfigure}%
	~
	\begin{subfigure}[t]{0.8\textwidth}
		\centering
		\includegraphics[width=13cm, height=15cm]{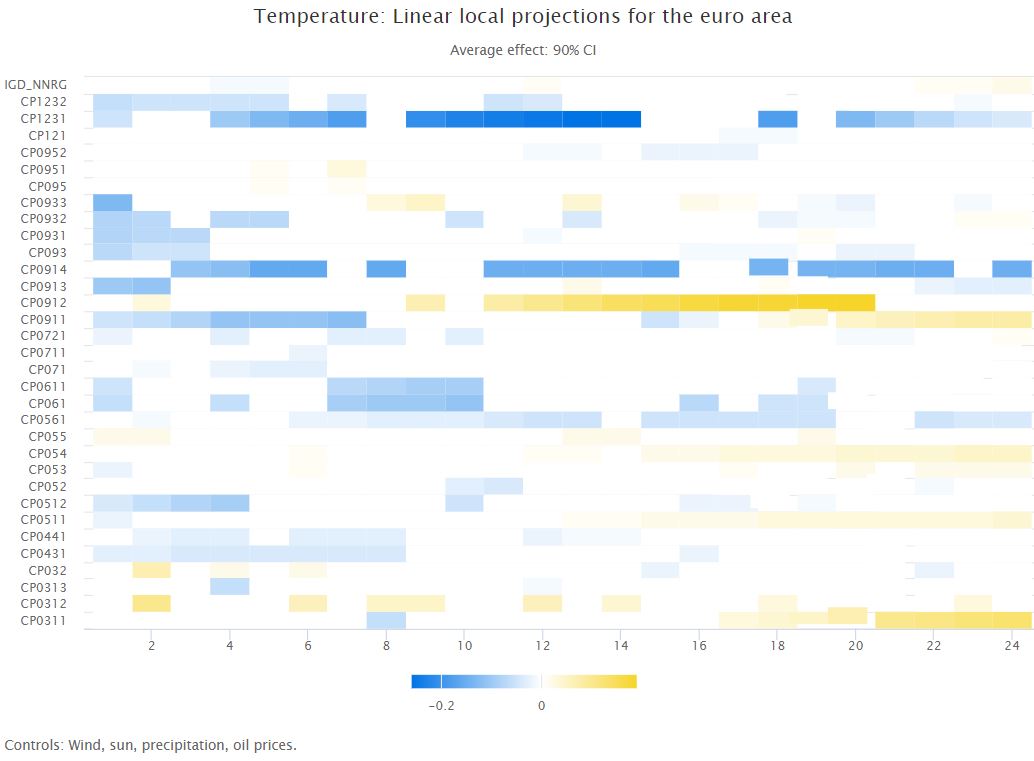}
	\end{subfigure}%
	\caption{Temperature: Local projections for the Euro Area (Non-Energy industrial goods)}		
		 \label{fig:FigA19}	
\end{figure*}	

\end{landscape}

\begin{landscape}
\captionsetup[subfigure]{aboveskip=1pt}
\begin{figure*}[th]
	\centering
	\begin{subfigure}[t]{0.8\textwidth}
		\centering
		\includegraphics[width=13cm, height=15cm]{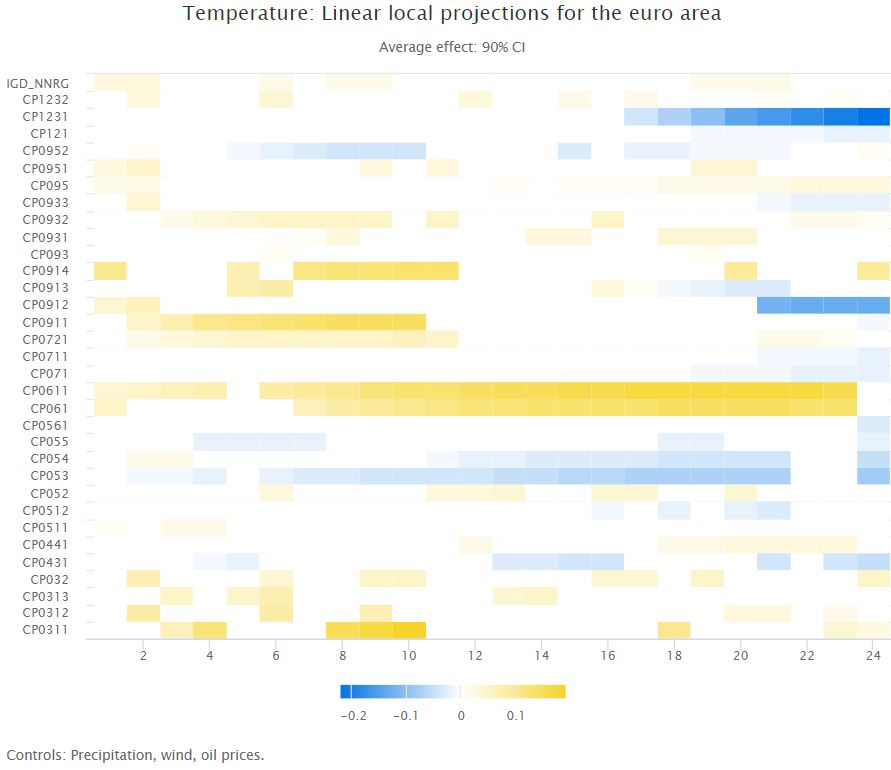}
	\end{subfigure}%
	~
	\begin{subfigure}[t]{0.8\textwidth}
		\centering
		\includegraphics[width=13cm, height=15cm]{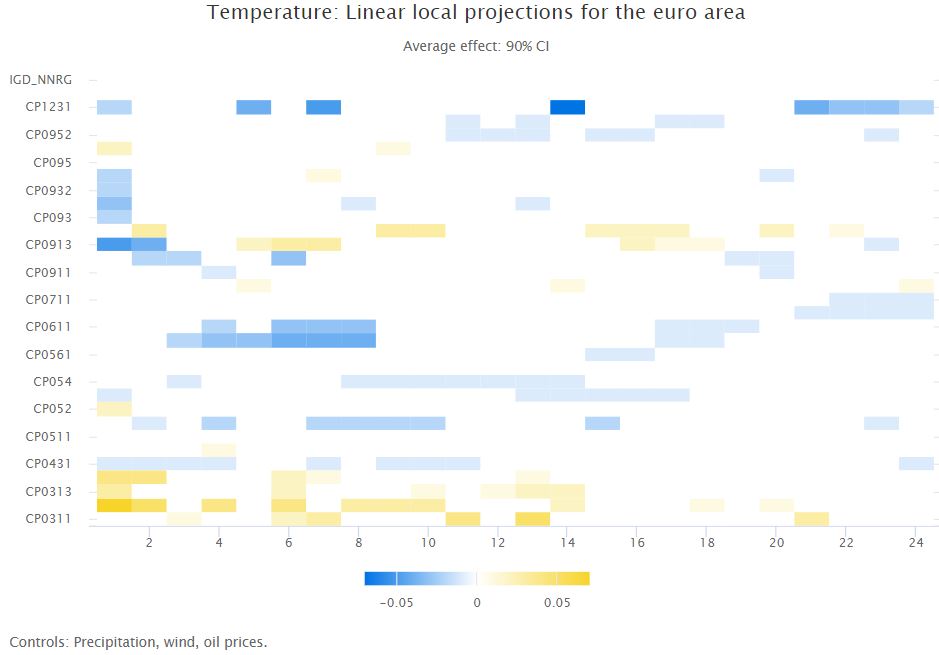}
	\end{subfigure}%
	\caption{Temperature: Local projections for the Euro Area (Non-Energy - Max and Min)}		
		 \label{fig:FigA20}	
\end{figure*}	
\end{landscape}

\begin{landscape}
\captionsetup[subfigure]{aboveskip=1pt}
\begin{figure*}[th]
	\centering
	\begin{subfigure}[t]{0.8\textwidth}
		\centering
		\includegraphics[width=13cm, height=15cm]{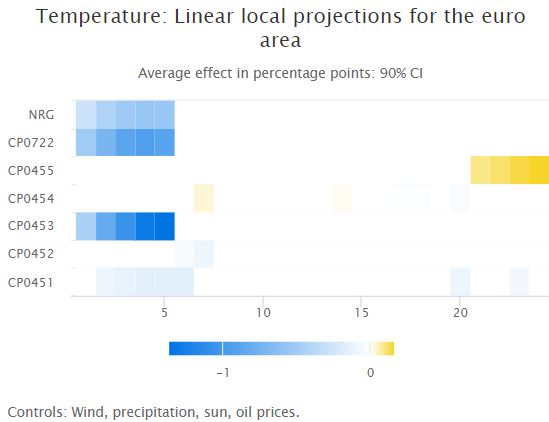}
	\end{subfigure}%
	~
	\begin{subfigure}[t]{0.8\textwidth}
		\centering
		\includegraphics[width=13cm, height=15cm]{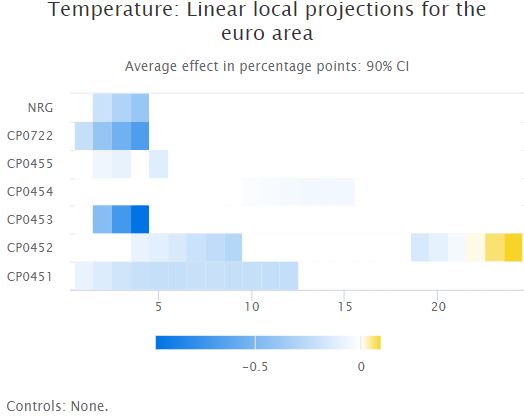}
	\end{subfigure}%
	\caption{Temperature: Local projections for the Euro Area (Energy)}		
		 \label{fig:FigA21}	
\end{figure*}	

\end{landscape}

\begin{landscape}
\captionsetup[subfigure]{aboveskip=1pt}
\begin{figure*}[th]
	\centering
	\begin{subfigure}[t]{0.8\textwidth}
		\centering
		\includegraphics[width=13cm, height=15cm]{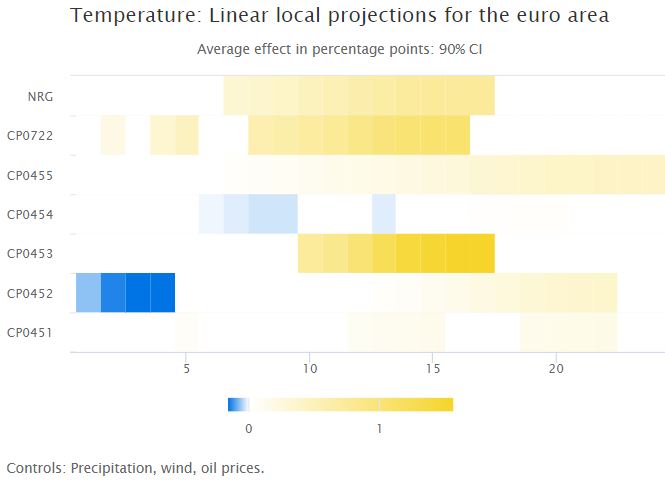}
	\end{subfigure}%
	~
	\begin{subfigure}[t]{0.8\textwidth}
		\centering
		\includegraphics[width=13cm, height=15cm]{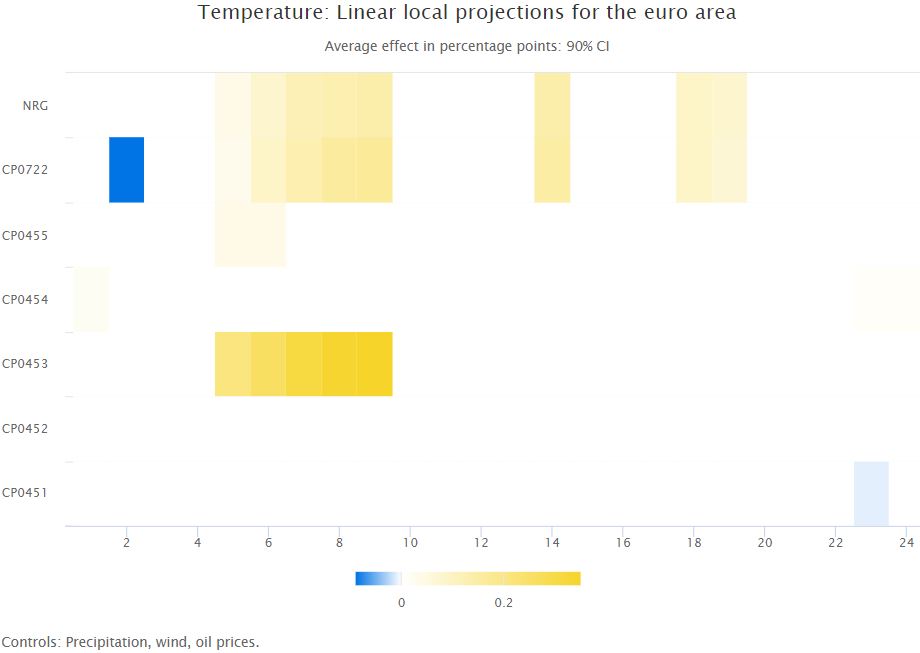}
	\end{subfigure}%
	\caption{Temperature: Local projections for the Euro Area (Energy - Max and Min)}		
		 \label{fig:FigA22}	
\end{figure*}	

\end{landscape}

\begin{landscape}
\captionsetup[subfigure]{aboveskip=1pt}
\begin{figure*}[th]
	\centering
	\begin{subfigure}[t]{0.8\textwidth}
		\centering
		\includegraphics[width=13cm, height=15cm]{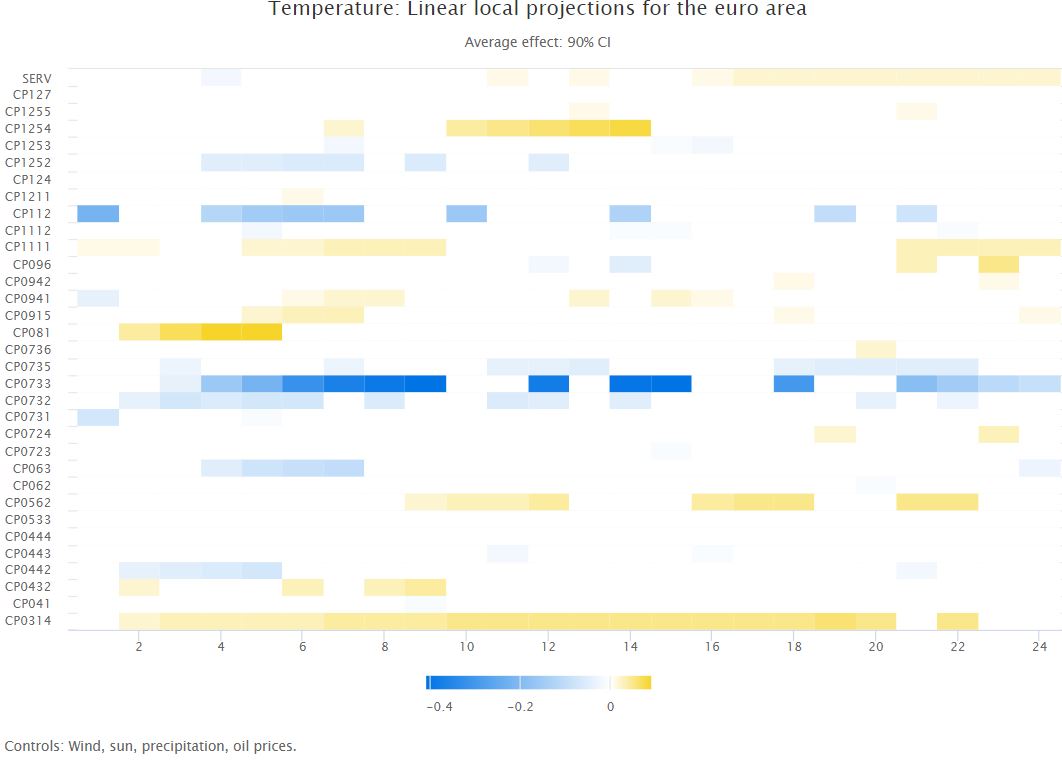}
	\end{subfigure}%
	~
	\begin{subfigure}[t]{0.8\textwidth}
		\centering
		\includegraphics[width=13cm, height=15cm]{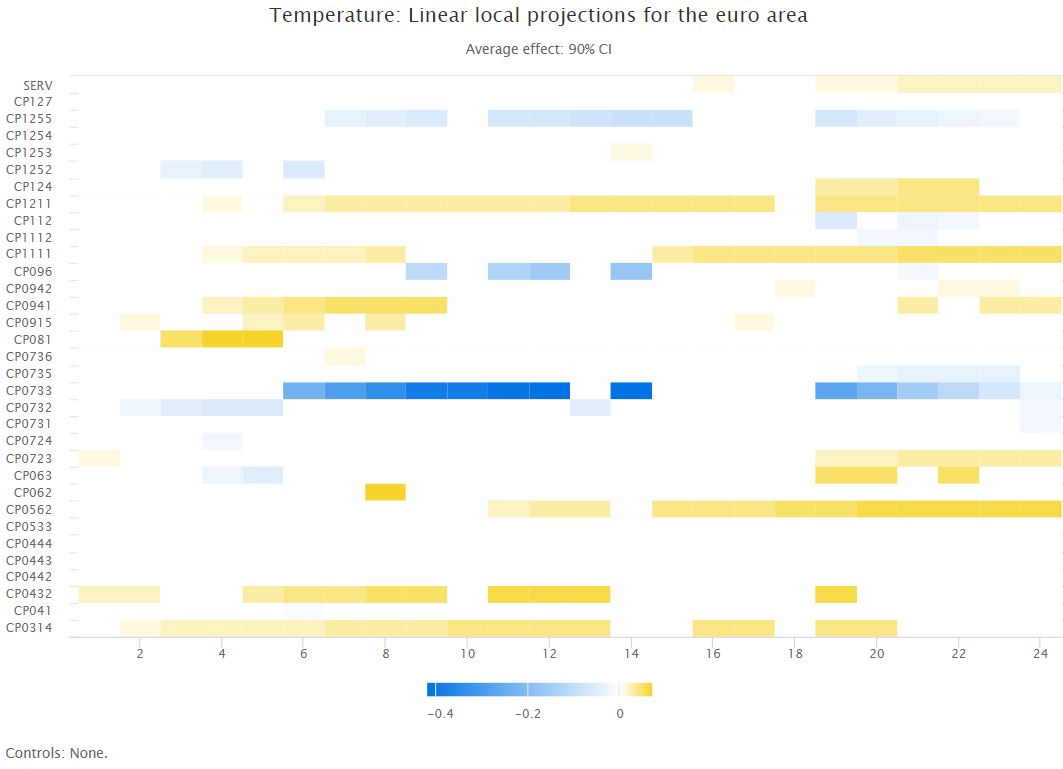}
	\end{subfigure}%
	\caption{Temperature: Local projections for the Euro Area (Services)}		
		 \label{fig:FigA23}	
\end{figure*}	

\end{landscape}

\begin{landscape}
\captionsetup[subfigure]{aboveskip=1pt}
\begin{figure*}[th]
	\centering
	\begin{subfigure}[t]{0.8\textwidth}
		\centering
		\includegraphics[width=13cm, height=15cm]{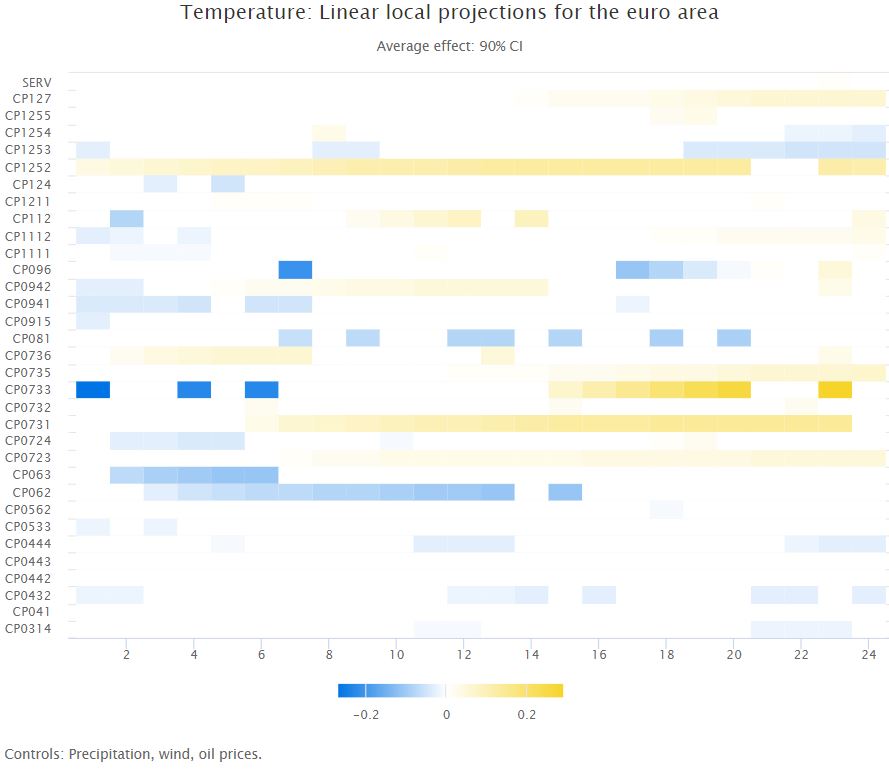}
	\end{subfigure}%
	~
	\begin{subfigure}[t]{0.8\textwidth}
		\centering
		\includegraphics[width=13cm, height=15cm]{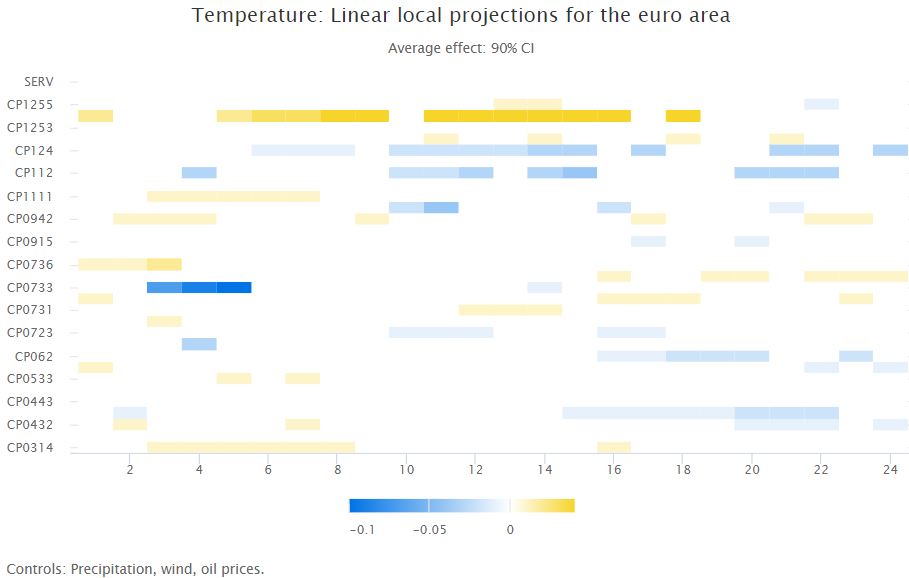}
	\end{subfigure}%

 	\caption{Temperature: Local projections for the Euro Area (Services). Left panel: Maximum temperature, right panel: Minimum temperature}		
		 \label{fig:FigA24}	
\end{figure*}	

\end{landscape}
\begin{landscape}
\begin{table}[bht!]
\caption{Average deviation from historical mean of different weather variables for the euro area countries, i.e., threshold values used above which the observed deviations are considered as shocks.}
\label{country_climate_deviations}
\begin{tabular}{lrrrrrr}
\toprule
Country & Temperature & Precipitation & Wind speed & Solar radiation & Minimum Temperature & Maximum Temperature\\
\midrule
AT & 1.40 & 0.06 & 1.11 & 2.93 & 1.32 & 1.42\\
BE & 1.39 & 0.02 & 1.99 & 7.44 & 1.15 & 1.82\\
CY & NA & NA & NA & NA & NA & NA\\
DE & 1.33 & -0.02 & 1.90 & 4.38 & 0.91 & 1.58\\
EE & 1.62 & 0.23 & 1.51 & 4.18 & 1.64 & 1.64\\
\addlinespace
EL & 0.25 & 0.04 & 1.33 & NA & 0.51 & 0.06\\
ES & 1.22 & -0.10 & 1.45 & NA & 0.93 & 1.55\\
FI & 1.55 & 0.20 & 1.59 & NA & 1.84 & 1.54\\
FR & 1.37 & 0.00 & 1.72 & 2.18 & 1.11 & 1.53\\
HR & 1.26 & 0.07 & 1.04 & 3.91 & 1.01 & 1.62\\
\addlinespace
IE & 0.74 & 0.09 & 2.04 & NA & 0.85 & 0.61\\
IT & 1.23 & -0.05 & 1.51 & -0.06 & 1.23 & 1.64\\
LT & 1.52 & 0.09 & 1.68 & 5.69 & 1.13 & 1.28\\
LU & 1.09 & -0.02 & 1.86 & 4.77 & 0.59 & 1.93\\
LV & 1.59 & 0.13 & 1.53 & 4.85 & 1.40 & 1.53\\
\addlinespace
MT & NA & NA & NA & NA & NA & NA\\
NL & 1.33 & 0.21 & 2.16 & 8.37 & 1.02 & 1.56\\
PT & 1.02 & -0.38 & 1.49 & NA & 1.07 & 1.37\\
SI & 1.29 & 0.10 & 0.60 & 5.64 & 1.16 & 1.73\\
SK & 1.35 & -0.07 & 1.29 & 4.18 & 1.38 & 1.28\\
\addlinespace
EA & 1.30 & 0.02 & 2.60 & 1.96 & 2.20 & 1.00\\
\bottomrule
\end{tabular}
\end{table}
\end{landscape}

\setcounter{equation}{2}

\begin{landscape}
\begin{table}[bht!]
\caption{Standard deviation of deviation from historical mean for euro area countries.}
\begin{tabular}{lrrrrrr}
\toprule
Country & Temperature & Precipitation & Wind speed & Solar radiation & Minimum Temperature & Maximum Temperature\\
\midrule
AT & 1.74 & 1.01 & 0.19 & 15.63 & 1.50 & 2.09\\
BE & 1.62 & 1.06 & 0.52 & 15.90 & 1.52 & 1.94\\
CY & NA & NA & NA & NA & NA & NA\\
DE & 1.70 & 0.82 & 0.41 & 15.51 & 1.53 & 2.05\\
EE & 2.38 & 0.83 & 0.37 & 14.94 & 2.65 & 2.28\\
\addlinespace
EL & 1.45 & 0.98 & 0.17 & NA & 1.35 & 1.56\\
ES & 1.23 & 0.72 & 0.24 & NA & 1.23 & 1.45\\
FI & 2.47 & 0.58 & 0.33 & NA & 2.78 & 2.29\\
FR & 1.50 & 0.91 & 0.36 & 14.25 & 1.37 & 1.79\\
HR & 1.73 & 1.41 & 0.18 & 15.44 & 1.56 & 2.02\\
\addlinespace
IE & 1.08 & 1.42 & 0.46 & NA & 1.14 & 1.18\\
IT & 1.20 & 1.08 & 0.19 & 13.29 & 1.19 & 1.37\\
LT & 2.27 & 0.77 & 0.35 & 15.37 & 2.46 & 2.31\\
LU & 1.67 & 1.11 & 0.48 & 16.91 & 1.55 & 2.07\\
LV & 2.31 & 0.74 & 0.29 & 15.16 & 2.57 & 2.27\\
\addlinespace
MT & NA & NA & NA & NA & NA & NA\\
NL & 1.59 & 1.02 & 0.57 & 14.73 & 1.53 & 1.85\\
PT & 1.18 & 1.54 & 0.28 & NA & 1.27 & 1.50\\
SI & 1.68 & 1.95 & 0.15 & 16.96 & 1.57 & 2.04\\
SK & 1.78 & 0.94 & 0.25 & 15.87 & 1.75 & 1.95\\
EA & 1.2 & 0.45 & 0.3 & 10.3 & & \\
\bottomrule
\end{tabular}
\end{table}
\end{landscape}

\setcounter{equation}{3}

\begin{table}[bht!]
\caption{Correlation of weather variables}
\begin{tabular}{lrl}
\toprule
 Weather variables & Correlation coefficient  \\
\midrule

Temperature \& Solar Radiation & 0.86 \\
Temperature \& Wind: Deviation & 0.34 \\
Temperature \& Solar Radiation: Deviation & 0.32 \\
Precipitation \& Wind: Deviation & 0.07 \\
Precipitation \& Wind & 0.05  \\
Wind \& Sun: Deviation & 0.01 \\
Temperature \& Wind & 0.00 \\
Wind \& Solar Radiation & -0.04 \\
Temperature \& Precipitation & -0.07 \\
Precipitation \& Solar Radiation & -0.26 \\
Precipitation \& Solar Radiation: Deviation & -0.62 \\

\bottomrule
\end{tabular}
\end{table}

\end{document}